\documentclass[prd,nofootinbib,floats,aps,twocolumn,tightenlines,superscriptaddress]{revtex4-1}

\usepackage{graphicx}
\usepackage{bbold}

\usepackage{hyperref}

\usepackage{bm}

\usepackage{amsfonts,amsmath,amssymb,amsthm}

\DeclareMathOperator{\Tr}{Tr}

\allowdisplaybreaks

\numberwithin{equation}{section}

\newcommand{\be}[1]{\begin{equation}\label{#1}}
\newcommand{\ee}{\end{equation}}

\newcommand{\ii}{\mathrm{i}}

\newcommand{\ket}[1]{\left| {#1} \right\rangle}
\newcommand{\bra}[1]{\left\langle {#1} \right|}

\newcommand{\braket}[2]{\left\langle {#1}\left|{#2}\right.\right\rangle}
\newcommand{\proj}[2]{\left| {#1} \right\rangle\!\left\langle {#2} \right|}

\newcommand{\sfx}{{\sf x}}

\newcommand{\BbbR}{\mathbb{R}}
\newcommand{\BbbZ}{\mathbb{Z}}
\newcommand{\BbbC}{\mathbb{C}}

\DeclareMathOperator{\Realpart}{Re}

\DeclareMathOperator{\erfc}{erfc}

\usepackage[usenames,dvipsnames]{xcolor}

\begin{abstract}
We study the impact of the zero mode of a quantum field 
on the evolution of a particle detector. 
For a massless scalar field in a periodic cavity, we show that the 
impact of the zero mode on the Unruh-DeWitt detector 
and its derivative-coupling generalisation is 
necessarily nonvanishing but can be made negligible in some limits, 
including those  commonly occurring  in non-relativistic quantum optics. 
For the derivative-coupling detector this can be accomplished by just tuning 
the zero mode's initial state, but the standard Unruh-DeWitt detector 
requires a more subtle  and careful tuning.  
Applications include an inertial detector with arbitrary velocity, 
where we demonstrate the regularity of the ultrarelativistic limit, 
and a detector with uniform acceleration. 
\end{abstract}

\begin{document}
\title{Particle detectors and the zero mode of a quantum field}

\author{Eduardo Mart\'in-Mart\'inez}
\affiliation{Institute for Quantum Computing, University of Waterloo, Waterloo, Ontario, N2L 3G1, Canada}
\affiliation{Department of Applied Mathematics, University of Waterloo, Waterloo, Ontario, N2L 3G1, Canada}
\affiliation{Perimeter Institute for Theoretical Physics, Waterloo, Ontario N2L 2Y5, Canada}
\author{Jorma Louko}
\affiliation{School of Mathematical Sciences, University of Nottingham, Nottingham NG7 2RD, UK}

\date{April 2014}

\maketitle

\section{Introduction\label{sec:intro}}

The response of a particle detector coupled to a quantum field has been subject of  extensive research since the 70s to  the present. 
While protocols that would directly measure the state of a quantum field are difficult to envisage, it is conceptually straightforward to make a measurement on a detector that has been allowed to interact with a quantum field. A spatially pointlike detector has a particular advantage in that it can be identified with an `observer' who is moving through the spacetime: 
such detectors have been used to quantify 
the particle content in a given state of a quantum field as seen 
 by a local observer \cite{Unruh1,DeWitt-inbook,Birrell1984,Wald1994,Crispino},
to analyze the entanglement contained in the vacuum state of a quantum field \cite{Reznik1}, to study metrology settings \cite{Dragan:2011zz}, to analyze the decoherence effects of relativistic trajectories \cite{lmt10,matsako}, to propose schemes of universal quantum computing via relativistic motion  \cite{AasenPRL} 
and to set up scenarios of quantum communication  in the relativistic limit \cite{Robort,Qcollcall}. 
The question `How many times  does a particle detector click for a given field state and a given trajectory in spacetime?' is relevant from quantum optics \cite{Scullybook} to the study of very fundamental problems as the quantum effects associated with the presence of  horizons  \cite{LeeThesis}, or to serve as a witness of primordial quantum fluctuations which may give information about the nature of the gravitational interaction~\cite{QuanG}.

To model the field-detector interaction it is commonplace to use the so-called 
Unruh-DeWitt (UDW) detector \cite{Unruh1,DeWitt-inbook}, 
which is a two-level quantum system that couples in a pointlike manner to a 
scalar field along its worldline. This model encompasses all the fundamental 
features of the light-matter interaction when there is no orbital angular momentum 
exchange involved \cite{Wavepackets,Alvaro}, 
and it is a useful tool for addressing a range of issues, from fundamental 
studies of the particle content in a given field state to quantum gravity, 
and from fundamental quantum optics to relativistic quantum  information. 
A limit of the UDW detector model yields the Jaynes-Cummings model, 
which is ubiquitous in quantum optics as a phenomenological model 
of light-matter interaction \cite{Scullybook}. 
Additionally, the UDW detector is a 
powerful effective model to describe the way in which 
superconducting qubits couple to microwave guides \cite{supercond,Wallraff04}. 



In this paper we  address the response of an UDW detector when the scalar field 
has a mode of vanishing frequency, known as a zero mode. Zero modes occur for 
a massless field in static cavities in 
Minkowski spacetime with Neumann or periodic boundary conditions, 
and they also occur in spatially compact cosmological spacetimes 
when the field is massless and couples conformally to the curvature~\cite{QuanG}. 
 Since the zero mode has a vanishing frequency, 
it cannot be treated as a harmonic oscillator, and it has no distinguished Fock vacuum.
In this paper we tackle the following question: 
How does the zero mode, and particularly the inherent ambiguity in its quantum state, 
affect the response of an UDW detector? 

 Eliminating the zero mode by hand from the field mode 
expansion would lead to an ill-defined quantum field theory, 
with nonvanishing commutators for the field operator at spacelike-separated events. 
Dropping the zero mode by hand from the coupling between 
the detector and the field  
would give a detector model that is as such mathematically consistent; however, 
as seen in~\cite{Alvaro}, 
the full linear coupling is 
necessary to model the $\bm p \cdot \bm A$ term by which 
an atomic electron couples to the quantized electromagnetic field. 
In quantum optics, these zero-mode issues arise with the common 
UDW and Jaynes-Cummings models with 
periodic and Neumann boundary conditions, and 
there it is usual to assume at the outset that any zero modes 
will have negligible effect and to drop them by hand. 
Our aim is to examine under which circumstances such dropping can be justified. 

We will see that a non-careful treatment of the zero mode is dangerous:  
the zero mode is able to produce important effects on the  detector's dynamics 
(both its click counting statistics and its quantum coherence). 
We will quantitatively study when these effects are maximized and minimized. 
In particular, we find that, by suitably choosing the initial state of the quantum field, 
it is indeed possible to minimize zero-mode effects on the detector dynamics under the UDW interaction. 
We will analyze  both the direct effects on the detector coming from 
the choice of the zero mode and the possible cross-talk effects coming 
from the effective coupling of the zero mode and 
the rest of the  field modes through their backreaction on the detector. 

We work in a periodic cavity in $(1+1)$-dimensional Minkowski spacetime. 
We consider both the usual UDW detector \cite{Unruh1,DeWitt-inbook}, which couples linearly to the scalar field, 
and its modification that couples linearly to the proper time derivative 
of the field~\cite{RavalHuAnglin,DaviesOttewill,Wang:2013lex,BenitoJorma}.

We start in Section \ref{sec:freefieldquant} by reviewing the 
quantisation of a massless scalar field in a $(1+1)$-dimensional periodic cavity, 
paying special attention to the zero mode and to its contribution to the two-point function. 
Section \ref{sec:densitymatrix2nd} analyses the density matrix of a conventional UDW detector 
and Section \ref{sec:dercoupling} the transition probability of UDW-type detector with a derivative coupling. 
The results are summarised and discussed in Section~\ref{sec:conclusions}. 

Throughout this paper, we use units in which $\hbar = c =1$. 
The spacetime signature is $(-+)$ 
where the minus direction is timelike.

\section{Preliminaries: 
massless scalar field in a periodic cavity\label{sec:freefieldquant}}

In this section we review the quantisation of a real, 
massless scalar field on a $(1+1)$-dimensonal flat, 
static spacetime with the spatial topology of a circle. 
In quantum optics terminology, this may be described as quantisation
in a periodic cavity, or as quantisation in $(1+1)$-dimensonal 
Minkowski spacetime with periodic boundary conditions. 
We shall establish the notation in a way that includes 
explicitly the zero-mode contributions to the mode expansion of the field 
and its two-point function. 

\subsection{Spacetime and quantum field}

The spacetime is a flat static cylinder with spatial circumference $L>0$. 
We work in standard Minkowski coordinates 
$(t,x)$ in which the metric reads 
\begin{align}
ds^2 = - dt^2 + dx^2
\ , 
\end{align}
with the periodic identification $(t,x) \sim (t,x+L)$. 

The quantum field is a free real massless scalar field $\phi$, 
with the action 
\begin{align}
S = \frac12 \int \text{d}t\, \text{d}x 
\left[ (\partial_t \phi)^2 - (\partial_x \phi)^2
\right]
\ , 
\label{eq:freefield-action}
\end{align}
and the field equation is $\square\phi=0$. The (indefinite) Klein-Gordon inner product reads 
\begin{align}
(\phi, \phi') = \ii \int \text{d}x \left( \phi^* \, \partial_t \phi' 
- \phi' \, \partial_t {\phi}^* \, \right) 
\ , 
\label{eq:KG-ip}
\end{align}
where the star denotes complex conjugation. 

We expand the Heisenberg picture field operator in the 
spatial Fourier modes in the usual fashion. We split the expansion as 
\begin{align}
\phi(t,x) = 
\phi_{\text{osc}}(t,x) 
+ 
\phi_{\text{zm}}(t) 
\ , 
\label{eq:phi-split}
\end{align}
where the oscillator-mode contribution $\phi_{\text{osc}}(t,x)$ contains 
the Fourier components that are not spatially constant 
and the zero-mode contribution $\phi_{\text{zm}}(t)$ contains the 
Fourier component that is spatially constant. 
We consider each contribution in turn.

\subsection{Oscillator modes\label{subsec:oscmodes}}

The positive frequency oscillator modes of the classical field are 
\begin{align}
\phi_n(t,x) = 
\frac{1}{\sqrt{4\pi |n|}}
\exp \! \left(-\ii \frac{2\pi |n|}{L} t  + \ii \frac{2\pi n}{L} x\right)
\ , 
\label{eq:phi-n-def}
\end{align}
where $n\in\BbbZ\setminus \{0\}$. The normalisation is such that 
$(\phi_m,\phi_n) = \delta_{mn}$. We thus have 
\begin{align}
\phi_{\text{osc}}(t,x) = 
\sum_{n\ne0} 
\bigl[
a_n \phi_n(t,x) + a_n^\dag \phi_n^*(t,x) 
\bigr]
\ , 
\label{eq:oscmodesexpansion}
\end{align}
where the nonvanishing commutators of the annihilation and creation operators 
are 
\begin{align}
[a_n , a^\dag_m] = \delta_{nm}
\ . 
\end{align}

The oscillator modes have a Fock vacuum $|0\rangle$, satisfying $a_n |0\rangle =0$. 
The oscillator-mode contribution to the Fock vacuum
Wightman function is 
\begin{align}
&\langle0|
\phi_{\text{osc}}(t,x)
\phi_{\text{osc}}(t',x')
|0\rangle
= 
\sum_{n\ne0}
\phi_n(t,x) 
\phi_n^*(t',x')
\notag
\\
& \ \ = 
\sum_{n=1}^{\infty}
\frac{1}{4\pi n}
\left\{
\exp \! \left[
-\ii \frac{2\pi n}{L} (\Delta u - \ii\epsilon)
\right]\right.
\notag
\\
&\ \ \qquad\qquad\left.+ 
\exp \! \left[
-\ii \frac{2\pi n}{L} (\Delta v -\ii\epsilon) 
\right]
\right\}
\ , 
\label{eq:oscmode-wightman}
\end{align}
where $u = t-x$, $v = t+x$, 
$\Delta u = u-u'$, $\Delta v = v-v'$, and $\epsilon\to0_+$. 
The sum in \eqref{eq:oscmode-wightman} can be evaluated in closed form, with the result 
\begin{align}
&\langle0|
\phi_{\text{osc}}(t,x)
\phi_{\text{osc}}(t',x')
|0\rangle
\notag
\\
&\ \ = 
- \frac{1}{4\pi}
\ln \left\{ 1 - \exp \! \left[
-\ii \frac{2\pi}{L} (\Delta u - \ii\epsilon)
\right] \right\}
\notag
\\
&\ \ \hspace{2.5ex}
- \frac{1}{4\pi}
\ln \left\{ 1 - \exp \! \left[
-\ii \frac{2\pi}{L} (\Delta v - \ii\epsilon)
\right] \right\}
\ . 
\label{eq:oscmode-wightman-summed}
\end{align}

\subsection{Zero mode\label{subsec:zm-quantisation}}

We denote the spatially constant Fourier component of the classical field by~$Q$. 
From \eqref{eq:freefield-action}, the 
Lagrangian for $Q$ 
is 
\begin{align}
L_{\text{zm}} = \frac{L}{2} {\dot Q}^2
\ . 
\end{align}
$Q$ has hence the dynamics of a nonrelativistic 
free particle on the real line, with $L$ taking the role of the mass. 
The Hamiltonian reads 
\begin{align}
H_{\text{zm}} = \frac{1}{2L} P^2
\ , 
\label{zmham}
\end{align}
where $P$ is the momentum conjugate to~$Q$. 

To quantise, let $Q_S$ and $P_S$ be the Schr\"odinger 
picture position and momentum operators corresponding to $Q$ and~$P$. 
$Q_S$ and $P_S$ are time-independent, they satisfy 
$[Q_S,P_S] = \ii$, and they commute with all $a_n$ and $a^\dag_n$. 
The Heisenberg picture position operator reads 
\begin{align}
Q_H(t) = Q_S + L^{-1}P_S t
\ . 
\label{eq:QH-def}
\end{align}

It follows that the zero-mode contribution to the 
Heisenberg picture field operator \eqref{eq:phi-split} is given by 
\begin{align}
\phi_{\text{zm}}(t) 
= 
Q_H(t)
\ , 
\label{eq:zeromode-expansion}
\end{align}
and $\phi_{\text{zm}}$ commutes with $\phi_{\text{osc}}$. 
While the zero mode does not have a Fock vacuum, its contribution 
to the Wightman function is nevertheless well defined: denoting the Heisenberg picture 
quantum state of the zero mode by $|\psi\rangle$, 
the zero-mode contribution to the 
Wightman function is
\begin{align}
\nonumber&\langle\psi|
\phi_{\text{zm}}(t) \phi_{\text{zm}}(t')
|\psi\rangle
= 
\langle\psi|
Q_S^2
|\psi\rangle
+ 
\langle\psi|
P_S Q_S
|\psi\rangle
L^{-1}t
\\&\quad+ 
\langle\psi|
Q_S P_S
|\psi\rangle
L^{-1}t'
+ 
\langle\psi|
P_S^2
|\psi\rangle
L^{-2} t t'
\ . 
\label{eq:zmode-wightman}
\end{align}

We emphasise that while the oscillator-mode contribution 
\eqref{eq:oscmode-wightman} to the Wightman function depends on 
$t$ and $t'$ only though the combination $t-t'$, 
the same does not hold for the 
zero-mode contribution \eqref{eq:zmode-wightman}: 
the Fock vacuum for the oscillator modes is time translation invariant, 
but the zero mode has no time translation invariant states. 
This will be significant in 
Sections \ref{sec:densitymatrix2nd} and \ref{sec:dercoupling} below. 

\subsection{Stress-energy tensor\label{subsec:stress-energy}}

The renormalised stress-energy tensor of the quantum field 
may be computed from the 
Wightman function by point-splitting~\cite{Birrell1984}, 
using as the short distance subtraction term the Minkowski 
vacuum Wightman function \cite{Birrell1984,PhysRevD.73.044027}, 
\begin{align}
&\langle
\phi(t,x)
\phi(t',x')
\rangle_{\text{Mink}}
= 
- \frac{1}{4\pi}
\ln \left[ (\epsilon + \ii \Delta u) (\epsilon + \ii \Delta v)
\right]
\ . 
\label{eq:mink-wightman}
\end{align}
We assume the field to be minimally coupled to curvature. 
When the oscillator modes are in the Fock vacuum, 
their contribution is \cite{Birrell1984}
\begin{align}
T_{tt}^{\text{osc}} = T_{xx}^{\text{osc}} = - \frac{\pi}{6 L^2}
\ , 
\ \ 
T_{tx}^{\text{osc}} = 0 
\ . 
\end{align}
The zero-mode contribution is 
\begin{align}
T_{tt}^{\text{zm}} = T_{xx}^{\text{zm}} 
= \frac{\langle\psi|
P_S^2
|\psi\rangle
}{2 L^2}
\ , 
\ \ 
T_{tx}^{\text{zm}} = 0 
\ .  
\label{eq:stress-energy-zm}
\end{align}

Note that the zero-mode contribution \eqref{eq:stress-energy-zm}
is time translation invariant, even though the Wightman function 
\eqref{eq:zmode-wightman} is not. 
Note also that both 
$T_{tt}^{\text{zm}}$ and $T_{xx}^{\text{zm}}$ are strictly positive.

\section{Density matrix of the UDW detector 
in second order perturbation theory\label{sec:densitymatrix2nd}}

In this section we consider the evolution of the UDW detector's 
reduced density matrix in second order perturbation theory in the coupling constant, 
identifying explicitly the contributions from the zero mode of the field. 
The full reduced density matrix, rather than just the transition probabilities, 
is required for examining for example how the detector 
suffers decoherence when interacting with an arbitrary state of the field. 

We would in particular like to identify situations 
where the effects of the zero mode on the detector's time evolution 
is negligible. This task consists of two steps: 
\begin{enumerate}
\item 
Identify those zero mode initial states (`safe' initial states) for which the 
zero mode has a negligible effect on the detector's evolution, assuming that 
back-reaction of the detector on the zero mode is neglected.  
\item 
Identify conditions under which a `safe' zero mode initial state remains 
`safe' under back-reaction from the detector. 
\end{enumerate}

We will see that guaranteeing the first condition is a very difficult endeavour. 
It will not be enough to demand that the 
zero mode initial state have vanishing energy, as 
one could have naively suspected.

On the other hand, we will see that the second condition can be satisfied 
by imposing constraints 
exclusively on the oscillator modes. If the oscillator 
modes are initialized in any ensemble of Fock states 
(Fock states or any diagonal density matrix in the Fock basis, 
such as a thermal state),  
a `safe' zero-mode state does remain `safe'.

\subsection{Coupled dynamics\label{subsec:coupled-dynamics}}

We consider a pointlike detector whose worldline $\sfx(\tau)$ 
is parametrised by the proper time~$\tau$. 
The detector is a two-level quantum system, 
with a Hilbert space spanned by the 
orthonormal energy eigenstates $\ket{g}$ and $\ket{e}$ 
whose respective energies are $0$ and~$\Omega$. 
If $\Omega>0$, $\ket{g}$ is the ground state and $\ket{e}$ is the excited state; 
if $\Omega<0$, the roles are reversed. We employ a notation that is adapted to the 
case $\Omega>0$, 
but all the formulas remain valid also for $\Omega<0$,  except in subsection 
\ref{subsec:zeromodeexample} where we take $\Omega>0$. 

The standard UDW interaction Hamiltonian is \cite{Unruh1,DeWitt-inbook,Birrell1984,Wald1994}
\begin{align}
H=\lambda \chi(\tau)\mu(\tau)\phi\bigl(\sfx(\tau)\bigr)
\ , 
\label{eq:H-int-def}
\end{align}
where $\mu(\tau)$ is the monopole moment operator, given by 
\begin{align}
\mu(\tau) = \sigma^+ e^{\ii\Omega\tau} + \sigma^- e^{-\ii\Omega\tau}
\ , 
\label{eq:monopole-def}
\end{align} 
and $\sigma^\pm$ are the usual raising and lowering operators, 
with the nonvanishing matrix elements 
$\bra{e} \sigma^+ \ket{g} = \bra{g} \sigma^- \ket{e} = 1$. 
The switching function $\chi$ specifies how the interaction is switched on and off. 
We assume that $\chi$ is smooth. We also assume either that $\chi$ 
has compact support, in which case 
the system is strictly uncoupled both before and after the interaction, 
or that $\chi$ has sufficiently strong falloff properties 
for the system to be treated as asymptotically uncoupled 
in the distant past and future. 

Working perturbatively to second order in~$\lambda$, 
the interaction picture time evolution operator $U$ 
for the full system is 
\begin{align}
U = U^{(0)} + U^{(1)} + U^{(2)} 
+ \mathcal{O}(\lambda^3)
\ , 
\label{eq:pert}
\end{align}
where 
\begin{subequations}
\begin{align}
U^{(0)} &= \openone
\ , 
\\
U^{(1)}
&= 
- \ii\int_{-\infty}^{\infty}\!\!\!\!\! \text{d} \tau 
\, H(\tau)
\ , 
\\
U^{(2)} &= 
- \int_{-\infty}^{\infty}
\!\!\!\!\!\text{d}\tau \!\!\int_{-\infty}^{\tau}\!\!\!\!\!\text{d}\tau'\, 
H(\tau) H(\tau')
\ . 
\label{eq:U2-def}
\end{align}
\end{subequations}
Given an initial density matrix $\rho_0$, the final density matrix 
$\rho_T$ is hence given by 
\begin{equation}
\rho_T=[\openone+U^{(1)}+U^{(2)}+\mathcal{O}(\lambda^3)]
\rho_0[\openone+U^{(1)}+U^{(2)}+\mathcal{O}(\lambda^3)]^{\dagger}
\ . 
\end{equation}
Writing $\rho_T=\rho_0+\rho_T^{(1)}+\rho_T^{(2)}+\mathcal{O}(\lambda^3)$, 
this gives 
\begin{subequations}
\begin{align}
\label{eq:rho1}\rho_T^{(1)}&=U^{(1)}\rho_0+\rho_0{U^{(1)}}^\dagger 
\ , 
\\
\label{eq:rho2}\rho_T^{(2)}&=U^{(1)}\rho_0{U^{(1)}}^\dagger+U^{(2)}\rho_0+\rho_0{U^{(2)}}^\dagger
\ . 
\end{align}
\end{subequations}

\subsection{Detector state: Evolution equations}

We take the detector's initial state to be an arbitrary 
density matrix, denoted by $\rho_{\text{d},0}$. 
For the initial state of the field, we assume 
that the zero mode is in a pure state~$\ket{\psi}$, 
and the oscillator modes are in a Fock state~$\ket{F}$, 
that is, in a pure state in which each of the modes is 
in a number operator eigenstate. 
The initial density matrix for the full coupled system is hence  
\begin{align}
\rho_0=\rho_{\text{d},0}\otimes\proj{F}{F}\otimes\proj{\psi}{\psi}
\ . 
\label{initrho}
\end{align}

The time evolution of the detector's density matrix is given by
\begin{align}
\rho_{d,T}=\text{Tr}_{\text{osc},\text{zm}}\left(U\rho_0U^\dagger\right)
\ , 
\end{align}
where the subscripts $\text{osc}$ and $\text{zm}$ 
indicate that the trace is over all the field degrees of freedom, 
including both the oscillator modes and the zero mode. 
We wish to analyse 
under what circumstances the contributions to $\rho_{d,T}$
can be separated into two decoupled parts, 
one accounting for the oscillator mode effects and 
the other for the zero mode effects. 

Recall first that the interaction Hamiltonian \eqref{eq:H-int-def} 
splits into the oscillator-mode contribution $H_{\text{osc}}$ 
and the zero-mode contribution 
$H_{\text{zm}}$ as 
\begin{subequations}
\label{eq:Hint-split}
\begin{align}
H&=H_{\text{osc}} +H_{\text{zm}}
\ , 
\\
H_{\text{osc}} & =\lambda \chi(\tau) \mu(\tau) 
\sum_{n\neq0} \bigl[ a_n \phi_n(x,t)  + a_n^\dagger \phi_n^*(x,t) \bigr]
\ , 
\\
H_{\text{zm}} & =
\lambda \chi(\tau)\mu(\tau)
\left(Q_S+\frac{P_S}{L}t\right)
\ , 
\label{zeroham}
\end{align}
\end{subequations}
$\phi_n$ is given by \eqref{eq:phi-n-def}, 
and $t$ and $x$ are understood as functions of $\tau$ since the field 
couples to the detector at the detector's location. It follows that 
\begin{subequations}
\label{eq:Uone-split}
\begin{align}
U^{(1)} 
& = U^{(1)}_{\text{osc}} + U^{(1)}_{\text{zm}}
\ , 
\\
U^{(1)}_{\text{osc}}&=-\ii\lambda\int_{-\infty}^\infty \!\!\!\text{d}\tau\,H_{\text{osc}}(\tau)
\ , 
\\
U^{(1)}_{\text{zm}}
& =
-\ii\lambda\int_{-\infty}^\infty\!\!\! \text{d}\tau\,H_{\text{zm}}(\tau)
\ . 
\end{align}
\end{subequations}
Writing 
\begin{align}
\rho_{0,\text{zm}}=\Tr_{\text{osc}} \rho_0
\ ,
\qquad \rho_{0,\text{d},{\text{osc}}}=\Tr_{\text{zm}} \rho_0
\ , 
\end{align}
we thus have 
\begin{align}
\Tr_{{\text{osc}},\text{zm}}\bigl({U^{(1)}\rho_0}\bigr)
& =
\Tr_{\text{osc}}{\bigl(U^{(1)}_{\text{osc}}\rho_{0,\text{d},{\text{osc}}}\bigr)}
\notag
\\
&\hspace{3ex}
+\Tr_{\text{zm}}{\bigl(U^{(1)}_{\text{zm}}\rho_{0,\text{d},\text{zm}}\bigr)}
\notag
\\
& =
\Tr_{\text{zm}}{\bigl(U^{(1)}_{\text{zm}}\rho_{0,\text{d},\text{zm}}\bigr)}
\ , 
\label{linear}
\end{align} 
where the last equality holds because 
$H_{\text{osc}}$ is off-diagonal in the Fock basis. 
From \eqref{eq:rho1} and \eqref{linear} 
we see that the order $\lambda$ contribution to  
$\rho_{d,T}$ comes entirely from the zero mode of the field. 

In order $\lambda^2$, the first term in \eqref{eq:rho2} gives 
\begin{align}
& \Tr_{{\text{osc}},\text{zm}}{\Bigl(U^{(1)}\rho_0{U^{(1)}}^\dagger\Bigr)}
\notag
\\
&\hspace{1ex}=
\Tr_{{\text{osc}},\text{zm}}{\Bigl(U^{(1)}_{\text{osc}}\rho_0{U_{\text{osc}}^{(1)}}^\dagger\Bigr)}
+\Tr_{{\text{osc}},\text{zm}}{\Bigl(U^{(1)}_\text{zm}\rho_0{U^{(1)}_\text{zm}}^\dagger\Bigr)}
\notag
\\
&\hspace{1ex}=
\Tr_{\text{osc}}{\Bigl(U^{(1)}_{\text{osc}}
\rho_{0,\text{d},{\text{osc}}}
{U_{\text{osc}}^{(1)}}^\dagger\Bigr)}
\notag
\\
&\hspace{4ex}
+\Tr_{\text{zm}}{\Bigl(U^{(1)}_\text{zm}
\rho_{0,\text{d},{\text{zm}}}
{U^{(1)}_\text{zm}}^\dagger\Bigr)}
\ , 
\label{mixed1}
\end{align}
where the cross terms that would involve both $U^{(1)}_{\text{osc}}$ 
and $U^{(1)}_\text{zm}$ 
are absent because $H_{\text{osc}}$ is off-diagonal in the Fock basis. 
For the last two terms in \eqref{eq:rho2}, 
using \eqref{eq:U2-def}, 
\eqref{eq:Hint-split} and \eqref{eq:Uone-split}
gives 
\begin{align}
\Tr_{{\text{osc}},\text{zm}}\bigl(U^{(2)}\rho_0\bigr)
& =
\Tr_{\text{osc}}\bigl(U^{(2)}_{\text{osc}}\rho_{0,\text{d},{\text{osc}}}\bigr)
\notag
\\
& \hspace{3ex}
+\Tr_{\text{zm}}\bigl(U^{(2)}_{\text{zm}}\rho_{0,\text{d},\text{zm}}\bigr)
\ , 
\label{ssss}
\end{align}
where
\begin{subequations}
\begin{align}
U^{(2)}_{\text{osc}}
&=- \int_{-\infty}^{\infty}\!\!\!\!\!\text{d}\tau \!\!
\int_{-\infty}^{\tau}\!\!\!\!\!\text{d}\tau'\, 
H_{\text{osc}}(\tau) H_{\text{osc}}(\tau')
\ , 
\\[1ex]
U^{(2)}_\text{zm}
& =
- \int_{-\infty}^{\infty}\!\!\!\!\!\text{d}\tau \!\!\int_{-\infty}^{\tau}
\!\!\!\!\!\text{d}\tau'\, 
H_\text{zm}(\tau) H_\text{zm}(\tau') 
\ , 
\end{align}
\end{subequations}
and terms involving $H_{\text{osc}}(\tau) H_\text{zm}(\tau')$ 
and $H_\text{zm}(\tau) H_{\text{osc}}(\tau')$ 
have vanished because $H_{\text{osc}}$ is off-diagonal in the Fock basis. 

Collecting, we see from \eqref{linear}, \eqref{mixed1} and 
\eqref{ssss}
that the detector's density matrix 
after the time evolution is given to order $\lambda^2$ by 
\begin{subequations}
\label{eq:rhoTd-coll}
\begin{align}
\rho_{T,\text{d}}
& =\rho_{0,\text{d}}
+{\rho^{\text{osc}}_{T,\text{d}}}^{(2)}
+{\rho^\text{zm}_{T,\text{d}}}^{(1)}
+{\rho^\text{zm}_{T,\text{d}}}^{(2)}
+ \mathcal{O}(\lambda^3)
\ , 
\\
{\rho^{\text{osc}}_{T,\text{d}}}^{(2)}
&=
\Tr_{\text{osc}}
{\Bigl(U^{(1)}_{\text{osc}}
\rho_{0,\text{d},{\text{osc}}}
{U_{\text{osc}}^{(1)}}^\dagger\Bigr)}
\notag
\\
&\hspace{3ex}
+ \Tr_{\text{osc}}\bigl(U^{(2)}_{\text{osc}}\rho_{0,\text{d},{\text{osc}}} 
+\text{H.c.}
\bigr)
\ , 
\label{eq:rhoTd-osc}
\\
{\rho^\text{zm}_{T,\text{d}}}^{(1)}
&= 
\Tr_{\text{zm}}{\bigl(U^{(1)}_{\text{zm}}\rho_{0,\text{d},\text{zm}}
+\text{H.c.}
\bigr)}
\ , 
\label{eq:rhoTd-zm1}
\\
{\rho^\text{zm}_{T,\text{d}}}^{(2)}
&= 
\Tr_{\text{zm}}{\Bigl(U^{(1)}_\text{zm}
\rho_{0,\text{d},{\text{zm}}}
{U^{(1)}_\text{zm}}^\dagger\Bigr)}
\notag
\\
&\hspace{3ex} 
+ \Tr_{\text{zm}}\bigl(U^{(2)}_{\text{zm}}\rho_{0,\text{d},\text{zm}}
+\text{H.c.}
\bigr)
\label{eq:rhoTd-zm2}
\ . 
\end{align}
\end{subequations}
The oscillator-mode contribution and the zero mode 
contribution to the detector's evolution 
are hence decoupled to order~$\lambda^2$: 
there is no `back-reaction' of the zero mode on the oscillator modes and 
vice versa through the interaction with the detector 
(the oscillator modes depositing energy in the detector 
and the detector transferring that energy into the zero mode).

We emphasise that this decoupling relies on the 
assumption that the oscillator modes are initially in a Fock state. 
The decoupling would continue to hold if the oscillator mode initial state were 
generalised to a diagonal density matrix in the Fock basis, 
such as for example a thermal state, 
but it would not hold for arbitrary initial states, 
such as, for instance, a coherent state.

\subsection{Detector state: Solution}\label{subsec:solution}

We now specialise to the case where 
the oscillator-mode initial state 
$\ket{F}$ is the Fock vacuum~$\ket{0}$. 

We employ a matrix representation in which 
\begin{align}
\ket{g}=\left (\!
\begin{array}{c}
1  \\
 0 \\
\end{array}
\!\right )
\ ,
\qquad \ket{e}=\left (\!
\begin{array}{c}
0  \\
1 \\
\end{array}
\!\right )
\ , 
\end{align}
so that the monopole moment operator 
\eqref{eq:monopole-def} has the form  
\begin{align}
\mu(\tau) = 
\left (\!
\begin{array}{c c}
0 & e^{-\ii \Omega \tau} \\
e^{\ii \Omega \tau} & 0 \\
\end{array}
\!\right )
\, . 
\label{eq:mu-matrixrep}
\end{align} 
We parametrise the detector's initial density matrix as 
\begin{align}
\rho_{0,\text{d}}=a\proj{g}{g}+b\proj{g}{e}+b^*\proj{e}{g}+(1-a)\proj{e}{e}
\ , 
\end{align}
where $a\in\BbbR$, $b\in\BbbC$, and 
${(a - \frac12)}^2 + {|b|}^2 \le \frac14$. 
The matrix representation is 
\begin{align}
\rho_{0,\text{d}}=\left (\!
\begin{array}{c c}
a & b \\
b^* & 1-a \\
\end{array}
\!\right )
\ . 
\label{eq:det-initialdmatrix-matrixform}
\end{align}

We address the oscillator-mode contribution and zero-mode contribution in 
\eqref{eq:rhoTd-coll} in turn. 


\subsubsection{Oscillator-mode contribution \eqref{eq:rhoTd-osc}}

To analyse the oscillator-mode contribution \eqref{eq:rhoTd-osc}, 
we introduce the quantities 
\begin{subequations}
\label{eq:omega-u-I-notation}
\begin{align}
u_n(\tau)
& =
\phi_n\bigl(t(\tau), x(\tau) \bigr)
\ , 
\\
\label{eq:omega-u-I-notation2}
I_{n,\pm}
& =
-\ii\lambda\int_{-\infty}^\infty \text{d}\tau\, 
\chi(\tau) \, e^{\pm \ii \Omega \tau} u^*_n(\tau)
\ , 
\\
G_{n,\pm}
&= 
 - \lambda^2 
\int_{-\infty}^\infty\!\!\!\! \text{d}\tau\int_{-\infty}^\tau\!\!\!\!
\text{d}\tau'\, \chi(\tau)\chi(\tau') \, e^{\pm \ii \Omega (\tau - \tau')}
\notag
\\
& 
\hspace{20ex}
\times 
u_n(\tau) u^*_n(\tau')
\ , 
\end{align}
\end{subequations}
where $n \in \BbbZ \setminus \{ 0\}$ and 
$\phi_n$ is the oscillator mode function \eqref{eq:phi-n-def}.  
In words, $u_n$ is the pull-back of $\phi_n$   
to the detector's worldline. 

Consider the first term in \eqref{eq:rhoTd-osc}. 
With the notation \eqref{eq:omega-u-I-notation}, 
we have 
\begin{align}
U_{\text{osc}}^{(1)} \rho_{0,\text{d},{\text{osc}}} 
&=\sum_{n\ne0}\left(I_{n,+} a_n^{\dagger}
\sigma^+ +I_{n,-} a_n^{\dagger}\sigma^-\right)
\rho_{0,\text{d},{\text{osc}}} 
\notag
\\
&=
\sum_{n\ne0}\proj{1_n}{0}\otimes
\Bigl[I_{n,+} \big(a \proj{e}{g}+b\proj{e}{e}\big)
\notag
\\
&\hspace{7ex}
+I_{n,-} \big(b^* \proj{g}{g}+(1-a)\proj{g}{e}\big)\Bigr]
\ . 
\end{align}
Multiplying from the right with ${U_{\text{osc}}^{(1)}}^\dagger$ 
and tracing over all the field modes, 
we find 
\begin{align}
&
\Tr_{\text{osc}}
{\Bigl(U^{(1)}_{\text{osc}}
\rho_{0,\text{d},{\text{osc}}}
{U_{\text{osc}}^{(1)}}^\dagger\Bigr)}
\notag
\\
& \hspace{2ex}
=
\sum_{n\ne0}\left (\!
\begin{array}{c c}
(1-a) {|I_{n,-}|}^2  & b^* I_{n,-} I_{n,+}^* \\[1ex]
b\, I_{n,-}^* I_{n,+} & a \, {|I_{n,+}|}^2 \\
\end{array}
\!\right )
\ . 
\label{eq:UnU}
\end{align}

Consider then the second term in \eqref{eq:rhoTd-osc}. 
We observe that 
\begin{align}
U^{(2)}_{\text{osc}}\rho_{0,\text{d},{\text{osc}}} 
&= 
 - \lambda^2 \sum_{n\ne0}\int_{-\infty}^{\infty}\!\!\!\!\text{d}\tau
\int_{-\infty}^{\tau}\!\!\!\!\text{d}\tau'\, 
\chi(\tau)\chi(\tau')\notag
\\
& \hspace{3ex}
\times u_n(\tau)u^*_n(\tau')
\, \mu(\tau)\mu(\tau') \, a_n a^\dagger_n
\, \rho_{0,\text{d},{\text{osc}}} 
\notag
\\
&\hspace{3ex}
+ \text{(traceless under $\Tr_{\text{osc}}$)}
\ . 
\end{align}
Writing out the product $\mu(\tau)\mu(\tau')$ using \eqref{eq:mu-matrixrep}, 
adding the Hermitian conjugate, and taking 
$\Tr_{\text{osc}}$, we obtain 
\begin{align}
&\Tr_{\text{osc}}\bigl(U^{(2)}_{\text{osc}}\rho_{0,\text{d},{\text{osc}}} 
+\text{H.c.}
\bigr)
\notag
\\
& \hspace{3ex}
= 
\sum_{n\ne0}
\left (\!
\begin{array}{c c}
2a\Realpart G_{n,-} & b \bigl(G_{n,-}+G^*_{n,+}\bigr) \\[1ex]
b^*\bigl(G_{n,+}+G^*_{n,-} \bigr) & 2(1-a) \Realpart G_{n,+} \\
\end{array}
\!\right )
\ , 
\label{eq:Un+nU}
\end{align}
where $G_{n,\pm}$ is given in \eqref{eq:omega-u-I-notation}. 

${\rho_{T,\text{d}}^{\text{osc}}}^{(2)}$ is hence given by the 
sum of \eqref{eq:UnU} and \eqref{eq:Un+nU}. 
As ${\rho_{T,\text{d}}^{\text{osc}}}^{(2)}$ 
must by construction be traceless and $a$ is a continuous parameter, 
setting the trace of this sum to zero yields the identities 
\begin{align}
0 = \sum_{n\ne0} 
\left(
{|I_{n,\pm}|}^2 + 2 \Realpart G_{n,\mp}
\right) 
\ , 
\end{align} 
which may be used to simplify ${\rho_{T,\text{d}}^{\text{osc}}}^{(2)}$ 
to the final form 
\begin{widetext}
\begin{align}
{\rho_{T,\text{d}}^{\text{osc}}}^{(2)}
&=
\sum_{n\ne0}
\left (\!
\begin{array}{c c}
(1-a) {|I_{n,-}|}^2 - a \, {|I_{n,+}|}^2
& b^* I_{n,-} I_{n,+}^* 
+ b \bigl(G_{n,-}+G^*_{n,+}\bigr) \\[1ex]
b\, I_{n,-}^* I_{n,+} + b^*\bigl(G_{n,+}+G^*_{n,-} \bigr) 
& a \, {|I_{n,+}|}^2 - (1-a) {|I_{n,-}|}^2 \\
\end{array}
\!\right )
\ . 
\label{eq:rho-osc2-f}
\end{align}


\subsubsection{Zero-mode contributions 
\eqref{eq:rhoTd-zm1} and \eqref{eq:rhoTd-zm2}}

To evaluate the order $\lambda$ zero-mode contribution \eqref{eq:rhoTd-zm1}, we note from 
\eqref{zeroham} that 
\begin{align}
U^{(1)}_{\text{zm}}\rho_{0,\text{d},\text{zm}}
= -\ii\lambda \int_{-\infty}^\infty\!\!\!\!\!\text{d}\tau\,\chi(\tau)\! 
\left( Q_S + \frac{P_S}{L} t(\tau) \right)\!\proj{\psi}{\psi} 
\left (\!
\begin{array}{c c}
0 & e^{-\ii \Omega \tau} \\
e^{\ii \Omega \tau} & 0 \\
\end{array}
\!\right )
\left (\!
\begin{array}{c c}
a & b \\
b^* & 1-a \\
\end{array}
\!\right )
\ . 
\end{align}
Adding the Hermitian conjugate and taking $\Tr_{\text{zm}}$, we find 
\begin{align}
{\rho^\text{zm}_{T,\text{d}}}^{(1)}
=
-\lambda \int_{-\infty}^\infty\!\!\!\!\text{d}\tau\,\chi(\tau)\left(  \langle Q_S\rangle_\psi+\frac{\langle P_S\rangle_\psi}{L} t(\tau) \right)
\left (\!
\begin{array}{c c}
2 \Realpart (\ii b^* e^{-\ii\Omega\tau}) & \ii(1-2a)e^{-\ii\Omega\tau} \\
 \ii(2a-1)e^{\ii\Omega\tau}& 2 \Realpart (\ii b e^{\ii\Omega\tau})  \\
\end{array}
\!\right )
\ . 
\label{firstzm}
\end{align}

The order $\lambda^2$ zero-mode contribution \eqref{eq:rhoTd-zm2} 
may be evaluated similarly, with the result 
\begin{align}
{\rho^\text{zm}_{T,\text{d}}}^{(2)}
& =
-\lambda^2 \int_{-\infty}^\infty\!\!\!\!\!\text{d}\tau\!\!
\int_{-\infty}^\infty\!\!\!\!\!\text{d}\tau'\,\chi(\tau)\chi(\tau')
\biggl( \langle Q_S^2\rangle_\psi 
+ \frac{\langle P_S^2\rangle_\psi}{L^2} t(\tau) t(\tau') 
+\frac{\langle Q_SP_S\rangle_\psi}{L} t(\tau') 
+\frac{\langle P_SQ_S\rangle_\psi}{L} t(\tau) \biggr)
\notag
\\[1ex]
& \hspace{20ex}
\times 
\left (\!
\begin{array}{c c}
(1-a)e^{-\ii \Omega (\tau-\tau')}& b^*e^{-\ii \Omega (\tau+\tau')} \\
b\,e^{\ii \Omega (\tau+\tau')} & a\,e^{\ii \Omega (\tau-\tau')} \\
\end{array}
\!\right )
\notag
\\[1ex]
& \hspace{3ex}
-\lambda^2\!\!\int_{-\infty}^\infty\!\!\!\!\!\text{d}\tau\!\!
\int_{-\infty}^\tau\!\!\!\!\!\text{d}\tau'\,\chi(\tau)\chi(\tau')
\biggl( \langle Q_S^2\rangle_\psi 
+ \frac{\langle P_S^2\rangle_\psi}{L^2} t(\tau) t(\tau')
+\frac{\langle Q_SP_S\rangle_\psi}{L} t(\tau')
+\frac{\langle P_SQ_S\rangle_\psi}{L} t(\tau) \biggr)
\notag
\\[1ex]
& \hspace{20ex}
\times 
\left (\!
\begin{array}{c c}
a\,e^{-\ii \Omega (\tau-\tau')}& b\,e^{-\ii \Omega (\tau-\tau')} \\
b^*e^{\ii \Omega (\tau-\tau')} & (1-a)e^{\ii \Omega (\tau-\tau')} \\
\end{array}
\!\right )
\notag
\\[1ex]
& \hspace{3ex}
-\lambda^2\!\!\int_{-\infty}^\infty\!\!\!\!\!\text{d}\tau\!\!
\int_{-\infty}^\tau\!\!\!\!\!\text{d}\tau'\,\chi(\tau)\chi(\tau')
\biggl( \langle Q_S^2\rangle_\psi 
+ \frac{\langle P_S^2\rangle_\psi}{L^2} t(\tau) t(\tau') 
+\frac{\langle P_SQ_S\rangle_\psi}{L} t(\tau') 
+\frac{\langle Q_SP_S\rangle_\psi}{L} t(\tau) \biggr)
\notag
\\[1ex]
& \hspace{20ex}
\times 
\left (\!
\begin{array}{c c}
a \, e^{\ii \Omega (\tau-\tau')}& b\,e^{-\ii \Omega (\tau-\tau')} \\
b^*e^{\ii \Omega (\tau-\tau')} & (1-a)e^{-\ii \Omega (\tau-\tau')} \\
\end{array}
\!\right )
\ . 
\label{secondzm}
\end{align}
\end{widetext}

\subsection{When is the zero-mode contribution small?\label{subsec:smallzeromode?}}

With the explicit formulas at hand, 
we are ready to address the central question: under what conditions are the 
zero-mode contributions to the detector's reduced density matrix small
compared with the oscillator-mode contributions? 

We can make the following observations. 

{\bf (i).}  
Suppose the zero mode initial state $\ket{\psi}$ is `safe', 
in the sense of negligible zero-mode contributions to the detector's evolution. 
As the oscillator-mode contributions 
and the zero-mode contributions to the detector's density matrix 
are decoupled, the total state remains `safe' throughout its evolution, 
to the order $\lambda^2$ in which we are working: 
the total state cannot become `unsafe' via interaction of the 
zero mode and the oscillator modes through their mutual interaction with the detector. 
We recall again that this 
conclusion relies on the oscillator modes having been prepared in a Fock state, 
and the conclusion would continue to hold also when the oscillator modes are prepared 
in a noncoherent ensemble of Fock states, such as a thermal state. 

{\bf (ii).} 
Suppose the zero mode initial state $\ket{\psi}$ has 
a nonvanishing expectation value of $Q_S$ or~$P_S$. 
Then the contribution of the zero mode 
to the detector's density matrix is not only non-negligible but in fact dominant: 
the zero mode gives a nonzero contribution already in order 
$\lambda$ while the oscillator-mode contributions appear only in order~$\lambda^2$. 

{\bf (iii).} 
Suppose the zero mode initial state $\ket{\psi}$ has 
vanishing expectation values of $Q_S$ and~$P_S$. 
This does not suffice to guarantee that the 
zero-mode contribution to the detector's evolution would be negligible: 
the zero-mode contributions occur then in the same $\lambda^2$ 
order as the oscillator-mode contributions, 
and they cannot be identically vanishing for any $\ket{\psi}$ 
as they are linear combinations of the expectation 
values of $Q_S^2$, $P_S^2$, $Q_S P_S$ and $P_S Q_S$. 

{\bf (iv).} 
Suppose the detector is initially in an  incoherent  superposition 
of the ground state and the excited state, so that $b=0$ in 
\eqref{eq:det-initialdmatrix-matrixform}. Then \eqref{firstzm} shows that the 
evolution of the diagonal elements in the 
detector's reduced density matrix is of order $\lambda^2$, 
regardless of the initial state of the zero mode. 
The zero mode can therefore not give the dominant 
perturbative contribution to the transition probabilities 
between the two eigenstates of the detector, 
although it can give the dominant contribution 
when the initial state is a coherent superposition of the two energy eigenstates. 

{\bf (v).} 
The explicit appearance of $t$ in 
\eqref{firstzm} and \eqref{secondzm} shows that the zero mode 
contribution to the detector's evolution is explicitly time-dependent. 
This is a consequence of the fact, noted in subsection \ref{subsec:zm-quantisation}, 
that the zero mode does not have time translation invariant states.

\subsection{Example: Zero mode in a 
harmonic oscillator ground state\label{subsec:zeromodeexample}} 

As an explicit example, we consider the zero mode state $\ket{\psi}$ 
whose wave function in the $Q$-representation is the Gaussian 
$\braket{Q}{\psi} =  {(\gamma/\pi)}^{1/4} \exp\bigl(- \frac12 \gamma Q^2\bigr)$ 
with $\gamma>0$. In words, 
$\ket{\psi}$ is a harmonic oscillator ground 
state with a frequency proportional to~$\gamma$.
We emphasise that as $\ket{\psi}$ is in the Heisenberg 
picture and the zero mode Hamiltonian 
$H_{\text{zm}}$ \eqref{zmham} is not that 
of a harmonic oscillator, 
the Schr\"odinger picture time evolution of $\ket{\psi}$ 
is not a pure phase but a spreading Gaussian. Nevertheless, 
in $\ket{\psi}$ we have 
\begin{subequations}
\begin{align}
\langle Q_S\rangle=\langle P_S\rangle=0
\ , 
\\
\langle Q_S^2\rangle
=\frac{1}{2\gamma} 
\ , 
\ \ \ 
\langle P_S^2\rangle
= \frac{\gamma}{2}
\ , 
\\
\langle Q_S P_S\rangle
= 
- \langle P_S Q_S\rangle
= 
\frac{\ii}{2} 
\ , 
\end{align}
\end{subequations}
and from \eqref{zmham} we further see that 
$\langle H_{\text{zm}}\rangle = \gamma/(4L)$. 
It follows from \eqref{firstzm}
that ${\rho^\text{zm}_{T,\text{d}}}^{(1)}=0$, 
and we can use \eqref{secondzm} 
to estimate ${\rho^\text{zm}_{T,\text{d}}}^{(2)}$. 
We can in particular ask whether the limit of small 
$\gamma$, in which the zero mode 
contribution to the stress-energy tensor 
\eqref{eq:stress-energy-zm}
is small, suffices 
to make also ${\rho^\text{zm}_{T,\text{d}}}^{(2)}$ small. 

We specialise to a detector that is static in the rest frame of the cylinder, therefore $t(\tau)=\tau$. 


To estimate the strength of the zero-mode contribution to the 
detector's dynamics, we introduce the following estimator modelled on the $U^{(1)}\rho_0U^{(1)}$ terms in~\eqref{secondzm}:
\begin{align} 
E^\pm_{\text{zm}}=&\frac{\lambda^2}{2}\!\!\int_{-\infty}^\infty\!\!\!\!\!\text{d}\tau\!\!\int_{-\infty}^\infty\!\!\!\!\!\text{d}\tau'\,\chi(\tau)\chi(\tau')
\nonumber \bigg[\! \frac{1}{\gamma}\!+\!\frac{\gamma}{2L^2}\tau\tau'\\
& +\frac{\ii}{L}(\tau'-\tau) \bigg]e^{\pm\ii \Omega (\tau-\tau')}
\ . 
\end{align}

For the Gaussian switching function
\begin{align}
\label{eq:gauss-switching}
\chi(\tau)  = 
\frac{1}{\pi^{1/4}\sigma^{1/2}}
e^{- \tau^2/(2\sigma^2)}
\ , 
\end{align}
where  the positive parameter $\sigma$ is the effective duration of the interaction, 
the integrals can be  evaluated analytically, 
yielding
%
\begin{align} 
E^\pm_{\text{zm}}&=\lambda^2\sqrt\pi e^{-\sigma ^2 \Omega ^2}\left(\frac{\gamma }{2L^2} \sigma ^5 \Omega ^2 +\frac{1}{\gamma }\sigma\mp \frac{2\sigma^3\Omega}{L}\right).
\label{estimzm}
\end{align}
Here, inspired by the usual quantum optics convention \cite{Scullybook}, we call $E^+_{\text{zm}}$ and $E^-_{\text{zm}}$ the counterrotating-wave  and rotating-wave contributions respectively. Note that the contributions from the zero mode of the counter-rotating wave terms $E^+_{\text{zm}}$ can be exactly cancelled by choosing $\gamma$ suitably. 
Specifically, for a given interaction time, length of the cavity and detector gap, we have two values of $\gamma$,
\be{gammazeros}\gamma_\pm=\frac{(2\pm\sqrt2 )L}{\sigma^2\Omega},\ee
which cancel the contribution to the detector dynamics coming from the zero-mode counter-rotating wave terms. The contribution of the zero-mode rotating-wave terms $E^-_\text{zm}$ cannot be cancelled for any value of $\gamma$, but we will discuss later that in this case its contribution to the detector dynamics is not relevant for the long interaction time regime.

To estimate the strength of the oscillator-mode contribution to the detector's dynamics, 
we introduce similar estimators
\begin{align} 
\label{estosc}
E_{\text{osc}}^\pm
=\sum_{n=1}^\infty 
\frac{\lambda^2}{2n\pi}\left|\int_{-\infty}^\infty \text{d}\tau\, 
\chi(\tau) \, e^{\ii(2\pi n/L \pm \Omega)\tau}\right|^2
\ , 
\end{align}
 modelled on the diagonal components of 
${\rho_{T,\text{d}}^{\text{osc}}}^{(2)}$~\eqref{eq:rho-osc2-f}.
For the Gaussian switching function \eqref{eq:gauss-switching}, we obtain 
\begin{align}
E_{\text{osc}}^\pm 
=
\sum_{n=1}^\infty
\frac{\lambda^2\sigma}{{n\sqrt{ \pi}}}
\exp \! \left[{-\frac{\sigma ^2 {(2 \pi  n \pm L \Omega)}^2}{ L^2}}\right]
\ . 
\end{align}
%

Let us study the relative strength of the zero-mode contribution as compared to the  oscillator-mode contribution
as a function of the zero mode natural frequency parameter $\gamma$ and the 
 effective total detection time~$\sigma$. The relevant relative strength estimator is
\begin{align} 
S^\pm_{\text{zm}}
& \equiv
\frac{|E_\text{zm}^\pm|}{| E^\pm_{\text{osc}}|}
\notag
\\
&=\pi e^{-\sigma ^2 \Omega ^2}\left|\frac{ \frac{\gamma }{2L^2} \sigma ^5 \Omega ^2 +\frac{1}{\gamma }\sigma\mp \frac{2\sigma^3\Omega}{L}}{\sum_{n=1}^\infty
\frac{\sigma}{{n}}
\exp \! \left[{-\frac{\sigma ^2 {(2 \pi  n \pm L \Omega)}^2}{ L^2}}\right]}\right|
\ . 
\label{eq:Spm-estimator}
\end{align}
 We may assume without loss of generality 
that $\Omega$ is positive, so that $\ket{g}$ is the ground state and $\ket{e}$ is the excited state  of the detector, 
as is the standard convention in quantum optics. 
Then $S^-_\text{zm}$ corresponds to the 
rotating-wave terms and $S^+_\text{zm}$ corresponds to the 
counter-rotating wave terms~\cite{Scullybook}. 
The behaviour for the two is markedly different.

Let us begin with the rotating-wave terms $S^-_\text{zm}$. 
We can see that if these terms contribute to the detector dynamics, and in the case where the detector is resonant with one of the field modes (i.e., if the gap of the atom coincides with an integer multiple of $2\pi/L$),  the impact of the resonant oscillator mode on the detector \eqref{estosc} will not be exponentially suppressed with the 
effective interaction duration~$\sigma$. 
Instead, the contribution of the resonant mode $2\pi n/L =\Omega$ will grow linearly with~$\sigma$, 
whereas the 
 contributions from the zero mode and from all the non-resonant oscillator modes are 
exponentially suppressed with~$\sigma$.

Looking at \eqref{eq:omega-u-I-notation2}, one can see that the terms $E^-_{\text{osc}}$ are nonexponentially vanishing only when we consider a detector which is excited and on resonance with one of the field modes, this is, when $\omega_n=\Omega$. It is very easy to see that similar terms would appear if we consider a detector  initially in its ground state but it is resonant with an  excited field mode (see for instance~\cite{Wavepackets}).  

This is a relieving result which rescues the usual quantum optical 
intuition behind the single-mode approximation~\cite{Scullybook}:  
In standard quantum optical setups, where we consider a detector's emission to or absorption from a
resonant field mode, the zero mode-dynamics can be safely neglected for large interaction times since it is exponentially suppressed with the duration of the 
interaction, whereas the  contribution of the resonant mode increases with time. 
By the same token, the contribution of the non-resonant modes can also be neglected for the stationary detector at rest. On the other hand, from \eqref{estimzm}, we also conclude that for a gapless detector $\Omega=0$, the zero-mode contribution will not be exponentially suppressed with the duration of the interaction. 
In this case, however, the effect of the zero mode can be made 
arbitrarily small by choosing $\gamma$ such that $\sigma/\gamma$ is small, 
as shown in~\eqref{estimzm}. 

\begin{figure*}[t]\begin{tabular}{cc}
\includegraphics[width=.43\textwidth]{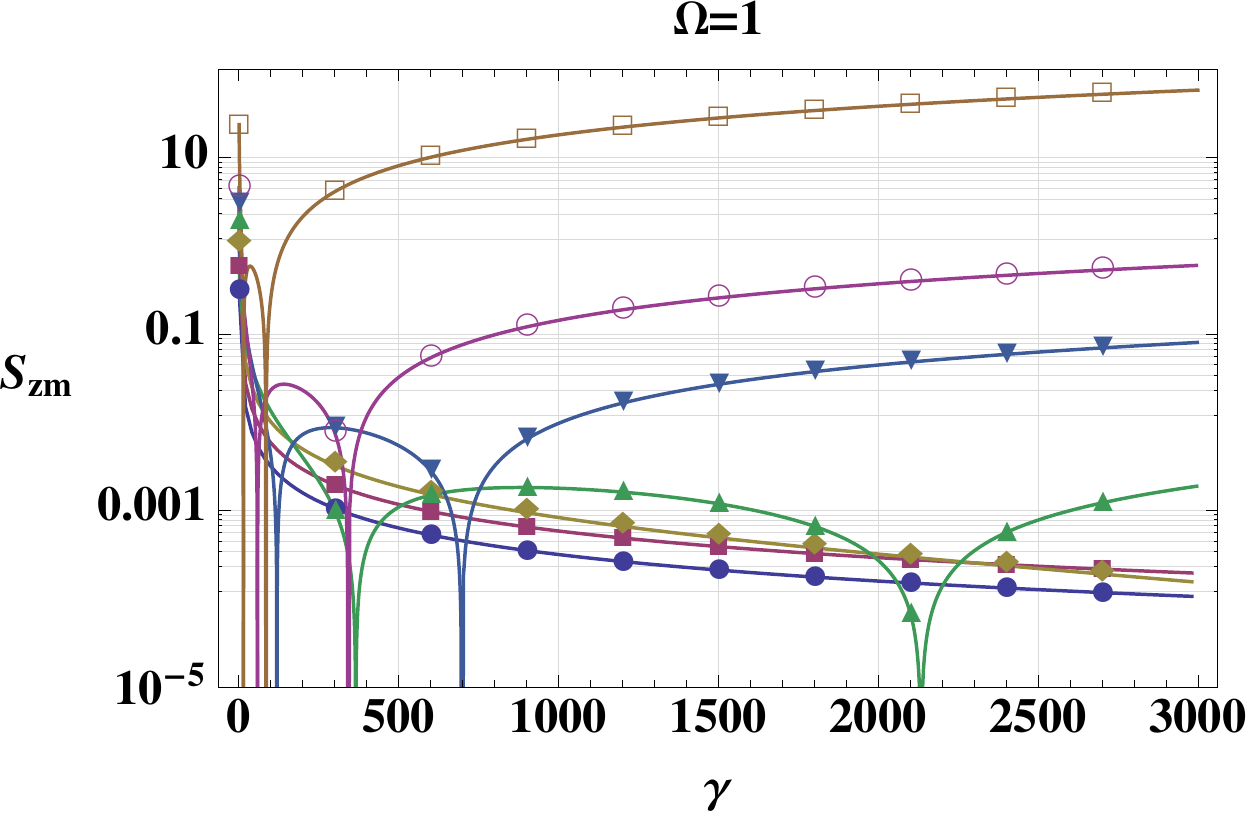}&
\includegraphics[width=.43\textwidth]{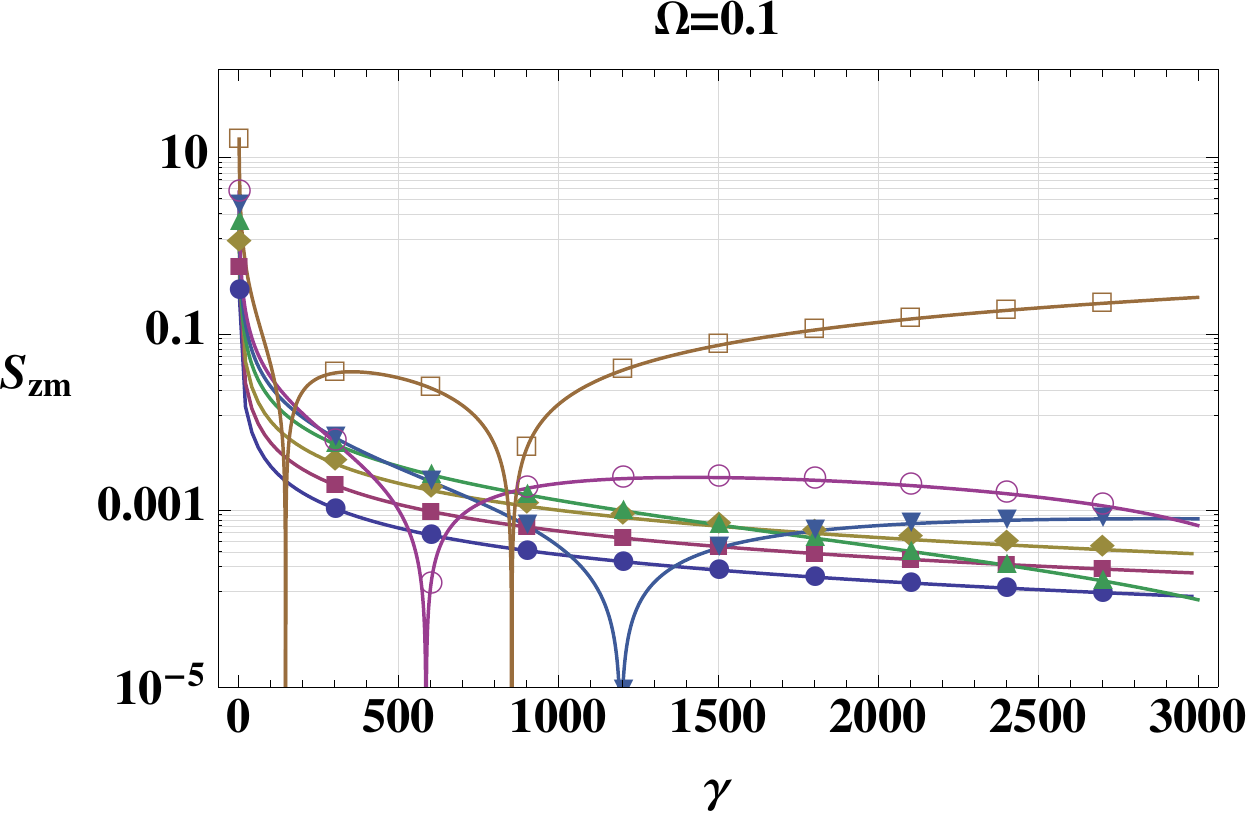}\end{tabular}
\caption{(Color online): The plots show the quantity $S^+_\text{zm}$ \eqref{eq:Spm-estimator}, 
which characterises the relative strength of the zero mode and oscillator mode 
contributions to the detector's density matrix in the counterrotating case, 
as a function of $\gamma$ for $\Omega=1$ and $\Omega=0.1$ with selected values of the effective interaction duration~$\sigma$. 
Solid circle is $\sigma=10^{-5}$, 
solid square is $\sigma=10^{-3}$, 
rhombus is $\sigma=10^{-2}$, 
triangle is $\sigma=0.04$, 
inverted triangle is $\sigma=0.07$, 
hollow circle is $\sigma=0.1$, 
and hollow square is $\sigma=0.2$. 
$S^+_\text{zm}$ has exact zeros at $\gamma_\pm=(2\pm\sqrt2 )L/(\sigma^2\Omega)$, 
which is within the plotted range of $\gamma$ only for the higher values of~$\sigma$.}
\label{estimator}
\end{figure*}

The counter-rotating contribution $S_\text{zm}^+$, however, is a whole different  story, 
as illustrated in Fig.~\ref{estimator}.
Notice that in this case the  impact  estimators of the zero mode and the oscillator modes on the detector state are both small when the effective interaction time $\sigma$ is long, 
but as shown in Fig.~\ref{estimator}, the contribution of the zero mode decays  with $\sigma$ much slower than the contribution of the oscillator modes. Although it is indeed possible to pick a value of $\gamma$ that cancels the relative strength of the zero mode in the detector's response,  this value would need to be large if the detector's energy gap is small. 
So the  zero-mode contribution can be exactly cancelled,  provided the value of $\gamma$ is adapted to~$\sigma$. 
As mentioned above,  from \eqref{estimzm}, we see that in the  gapless detector limit $\Omega\rightarrow0$, the contribution of the zero mode to the dynamics can be made  arbitrarily small by taking $\sigma/\gamma$ sufficiently small, 
but this will not be the case for non-zero detector gaps, where the value of $\gamma$ such that the contribution of the zero mode exactly cancels is a function of~$\sigma$. 

Nevertheless, if the gap of the detector increases, we see that the relative strength of the zero mode as compared with the impact of the oscillator modes becomes non-negligible if the interaction timescale is large enough, and even gives the leading contribution in the limit of very long interaction times.

This is hinting  that the impact of the zero mode on the detector dynamics, while minimizable by choosing appropriate initial states for the zero mode, may be highly non-negligible for large gap detectors and for long interaction times,  provided that there are no resonance-effects of the atom capturing or emitting  quanta into a resonant mode, as  it is for example the case of a detector in the ground state coupled to the field vacuum.

Additionally, one may argue that it could be challenging to ever experimentally acknowledge these effects by probing the state of the detector: the zero mode becomes more relevant for long times, but in absence of spontaneous emission or absorption, the vacuum response of the detector for both the zero mode and the oscillator modes is exponentially suppressed with time. Indeed, an inertial UDW detector of positive gap in the ground state, which remained on forever while interacting with the vacuum, has a vanishing probability of excitation, as it can be checked from \eqref{eq:rho-osc2-f} considering $\Omega>0$. This means that even though the zero-mode contribution is relatively large as compared to the oscillator-mode contribution,  in the regimes where this happens the overall action on the detector would be negligible.  Notice that this may be amplified  by choosing a more sudden switching than a Gaussian smearing. 

We also make the following hypothesis: 
Another scenario where these effects may be non-trivial is in 
the interaction of more than one detector with the same quantum field, 
and in the extraction of correlations from the background field \cite{Reznik1,farming}.

\section{Transition probability of a 
derivative-coupling detector\label{sec:dercoupling}}

\subsection{Derivative-coupling detector}

We have seen that the zero-mode contribution to the evolution 
of the UDW detector is well defined, and in certain circumstances it can  
be arranged to be small compared with the oscillator-mode contribution. 
One unappealing property of the zero-mode contribution however 
is that it is not invariant under time translations in any state of the zero mode, 
not even when the detector is stationary. 
Further, the terms involving $t(\tau)$ and $t(\tau')$ in \eqref{firstzm} and \eqref{secondzm} 
show that the time dependence may be significant, 
with potentially polynomial growth at late times. 

In this section we consider a modified UDW detector 
for which the zero-mode contribution to the detector's transition probabilities 
is time translation invariant. 
In the notation of Section~\ref{sec:densitymatrix2nd}, 
the new interaction Hamiltonian is 
\begin{align}
H=\lambda \chi(\tau)\mu(\tau){\dot\phi}\bigl(\sfx(\tau)\bigr)
\ , 
\label{eq:H-int-dc-def}
\end{align}
where the overdot denotes derivative with respect to~$\tau$, 
 and the assumptions on the switching function $\chi$ are as in 
Section~\ref{sec:densitymatrix2nd}. 
In words, the detector couples to the proper 
time derivative of the field at the detector's location, rather than
to the field itself. 
The derivative ameliorates the
effects that stem from the 
infrared behaviour of a massles field in 
$(1+1)$ dimensions in a number of 
contexts \cite{RavalHuAnglin,DaviesOttewill,Wang:2013lex,BenitoJorma}, 
and in our context it will 
restore time translation invariance. 

A price to pay for the derivative in \eqref{eq:H-int-dc-def} is, however,  
that the detector's ultraviolet properties become similar to those  
of the non-derivative UDW detector \eqref{eq:H-int-def} in $(3+1)$ dimensions. 
If we try to proceed with \eqref{eq:H-int-dc-def}
as in Section~\ref{sec:densitymatrix2nd}, 
we find that the off-diagonal components in the counterpart of 
${\rho_{T,\text{d}}^{\text{osc}}}^{(2)}$ 
\eqref{eq:rho-osc2-f} are ill-defined, 
due to a divergence of the sums at large~$|n|$. 
This ultraviolet problem occurs even in the simpler setting of a  
non-derivative UDW detector \eqref{eq:H-int-def} 
in $(3+1)$ dimensional Minkowski spacetime, 
as can be seen by expressing the time evolution of the detector's density 
matrix in terms of
the two-point function of the field as in \cite{Robort}
and considering the short-distance Hadamard form of the
two-point function~\cite{PhysRevD.73.044027,Kay1991}. 

We shall therefore not consider the detector's full density 
matrix but just the diagonal elements, 
which give the transition probabilities. 
Adapting the standard treatment of the non-derivative 
UDW detector \cite{Unruh1,Birrell1984,DeWitt-inbook,Wald1994} 
and working to leading order in~$\lambda$, 
the probability of a transition from $\ket{g}$ to $\ket{e}$ 
is equal to 
\begin{align}
P(\Omega) = 
\lambda^2 
F(\Omega)
\ , 
\label{eq:probability-der-def}
\end{align}
where the response function $F$ is given by 
\begin{align}
F(\Omega)
= 
\int \text{d}\tau \, \text{d}\tau' 
\, 
\chi(\tau) \chi(\tau') 
\, 
e^{-\ii\Omega (\tau-\tau')}
\partial_\tau \partial_{\tau'}
W(\tau,\tau')
\ , 
\label{eq:respfunction-der-def}
\end{align}
and the correlation function $W$ 
is obtained by pulling back the field's Wightman function, 
in the unperturbed state of the field, to the detector's worldline. 
Note that $W$ is a distribution and must be given an appropriate 
$\ii\epsilon$ prescription. For us this prescription will be straightforward, 
but it would require more attention if one wished to address the
limit in which the interaction is 
switched on and off 
 sharply~\cite{BenitoJorma,Schlicht2004,Langlois2006,Satz2006,Satz2007,Louko:2007mu,Hodgkinsonclick,Hodgkinson2012}. 

From now on we drop the factor $\lambda^2$ in \eqref{eq:probability-der-def} 
and refer to the response function \eqref{eq:respfunction-der-def} as the probability. 

We specialise to the situation in which the field is initially 
in the state discussed in Section \ref{sec:freefieldquant}: 
in the free field Heisenberg picture, 
the oscillator modes are in the Fock vacuum $\ket{0}$ 
and the zero mode is in a state~$\ket\psi$. 
The Wightman function is then given by the sum of the oscillator-mode contribution 
\eqref{eq:oscmode-wightman}
and the zero-mode contribution~\eqref{eq:zmode-wightman}.

\subsection{Zero-mode contribution to the response function\label{subsec:der-smallzeromode?}}

By \eqref{eq:zmode-wightman} and 
\eqref{eq:respfunction-der-def}, the zero-mode contribution to the response function reads 
\begin{align}
F_{\text{zm}}(\Omega)
= 
\frac{\langle\psi|
P_S^2
|\psi\rangle}{L^2}
\left|
\int_{-\infty}^\infty \text{d}\tau \, 
\chi(\tau) \, e^{-\ii \Omega \tau} \frac{dt(\tau)}{d\tau} \right|^2
\ . 
\label{eq:der-zmode-response-gen}
\end{align}
Two crucial observations are immediate. 

First, since $F_{\text{zm}}$ depends on $t(\tau)$ only through its derivative, 
any additive constant in $t(\tau)$ will drop out. $F_{\text{zm}}$
is manifestly invariant under time translations. 

Second, $F_{\text{zm}}$ depends on $\ket{\psi}$ only in that 
it is proportional to the single expectation value $\langle\psi| P_S^2 |\psi\rangle$. 
As $P_S^2$ is a positive definite operator with a continuous spectrum, 
$\langle\psi| P_S^2 |\psi\rangle$
is strictly positive for any $\ket{\psi}$; 
however, $\langle\psi| P_S^2 |\psi\rangle$
can be made arbitrarily small by a suitable choice of~$\ket{\psi}$. 
In this sense, $F_{\text{zm}}$ can always be made as small 
as desired by a suitable choice of the state of the zero mode. 

We emphasise that both of these properties are in a marked 
contrast with those of the non-derivative 
detector of Section~\ref{sec:densitymatrix2nd}. 
In \eqref{secondzm}, additive constants in $t(\tau)$ do not drop out, 
and the matrix elements involving $\ket{\psi}$ cannot all be 
simultaneously made arbitrarily small for any choice of~$\ket{\psi}$.  We also note that the $\ket{\psi}$-dependent overall coefficient 
$\langle\psi| P_S^2 |\psi\rangle / L^2$ 
in \eqref{eq:der-zmode-response-gen} is, 
up to a numerical factor, equal to the $tt$ and $xx$ 
components of the zero-mode contribution to the field's 
stress-energy tensor~\eqref{eq:stress-energy-zm}.

\subsection{Stationary detector}

\subsubsection{Response of a stationary detector}

As a first example, we consider a detector on the inertial worldline 
\begin{align}
t = \tau \cosh\beta 
\ , 
\ \ \ \ 
x = \tau \sinh\beta
\ , 
\label{param}
\end{align}
where $\beta\in\BbbR$ is the rapidity with respect to the
cylinder's rest frame, so that the 
detector's velocity with
respect to the cylinder's rest frame is $\tanh\beta$. 
This is the most general stationary trajectory on the cylinder. 

Pulling back \eqref{eq:oscmode-wightman} and \eqref{eq:zmode-wightman}
to the worldline~\eqref{param}, 
we find that the correlation function is given by 
\be{eq12}W(\tau,\tau') 
= 
W_{\text{osc}}(\tau,\tau') 
+ 
W_{\text{zm}}(\tau,\tau'),
\ee
where
\begin{widetext}
\begin{subequations}
\begin{align}
W_{\text{osc}}(\tau,\tau') 
&= 
\langle0|
\phi_{\text{osc}}\bigl(t(\tau),x(\tau)\bigr)
\phi_{\text{osc}}\bigl(t(\tau'),x(\tau')\bigr)
|0\rangle
\!=\!\! 
\sum_{n=1}^{\infty}
\frac{1}{4\pi n}
\left\{
\exp \! \left[
 \frac{2\pi n e^{-\beta}}{\ii L} (\Delta \tau - \ii\epsilon)
\right]
+ 
\exp \! \left[
\frac{2\pi n e^{\beta}}{\ii L} (\Delta \tau -\ii\epsilon) 
\right]
\right\}
\ , 
\label{eq:oscmode-corr}
\\[1ex]
W_{\text{zm}}(\tau,\tau') 
&= 
\langle\psi|
Q_H\bigl(t(\tau)\bigr) Q_H\bigl(t(\tau')\bigr)
|\psi\rangle
\notag
\\
&= 
\langle\psi|
Q_S^2
|\psi\rangle
+ 
\langle\psi|
P_S Q_S
|\psi\rangle
L^{-1}\tau \cosh\beta 
+ 
\langle\psi|
Q_S P_S
|\psi\rangle
L^{-1}\tau' \cosh\beta 
+ 
\langle\psi|
P_S^2
|\psi\rangle
L^{-2} \tau \tau' \cosh^2 \! \beta 
\ , 
\label{eq:zmode-corr}
\end{align}
\end{subequations}
where $\Delta\tau = \tau - \tau'$. 
Hence 
\begin{subequations}
\label{eq:bothmodes-corr-dd}
\begin{align}
\partial_\tau \partial_{\tau'}
W_{\text{osc}}(\tau,\tau') 
&= 
\frac{\pi}{L^2}
\sum_{\eta=\pm1}
\sum_{n=1}^{\infty}
n e^{-2\eta\beta}
\exp \! \left[
-\ii \frac{2\pi n e^{-\eta\beta}}{L} (\Delta \tau - \ii\epsilon)
\right]
\label{eq:oscmode-corr-dd}
\ , 
\\
\partial_\tau \partial_{\tau'}
W_{\text{zm}}(\tau,\tau') 
&= 
\frac{\cosh^2 \! \beta }{L^2}
\langle\psi|
P_S^2
|\psi\rangle
\ , 
\label{eq:zmode-corr-dd}
\end{align}
\end{subequations}
\end{widetext}
where $\eta=1$ indicates terms that come from the right-movers (field modes with positive momentum)  
and $\eta=-1$ indicates terms that come from the left-movers (field modes with negative momentum). 
From \eqref{eq:respfunction-der-def} we then obtain 
\begin{subequations}
\label{eq:bothmodes-response}
\begin{align}
F(\Omega) 
&= 
F_{\text{osc}}(\Omega) + F_{\text{zm}}(\Omega)
\ , 
\\
F_{\text{osc}}(\Omega)
&= 
\sum_{\eta=\pm1}
\sum_{n=1}^{\infty}
\left|I_n^\eta\right|^2
\ , 
\\
I_n^\eta
&= 
\frac{\sqrt{\pi n}}{L}
e^{-\eta\beta}
{\hat\chi}\left( \Omega + \frac{2\pi n e^{-\eta\beta}}{L}\right) 
\ , 
\\
F_{\text{zm}}(\Omega)
&= 
\frac{\cosh^2 \! \beta}{L^2}
\langle\psi|
P_S^2
|\psi\rangle
\left|\hat\chi(\Omega)\right|^2
\ , 
\label{eq:zmode-response}
\end{align}
\end{subequations}
where 
$\hat\chi$ is the Fourier transform of~$\chi$, 
\begin{align}
\hat\chi(\omega)  = \int_{-\infty}^\infty \text{d}\tau \, 
e^{-\ii\omega\tau} \chi(\tau)
\ . 
\end{align}

We note that the detector's velocity enters $F_{\text{osc}}$ 
through a Doppler shift: 
the field modes with momentum in (respectively opposite to) 
the detector's velocity contribute with a redshift (blueshift).

\subsubsection{Limits of long detection and ultrarelativistic velocity\label{subsubsec:der-limits}}

We wish to examine the response in the limit of long detection 
and in the limit of ultrarelativistic velocity. 
We choose the switching to be the Gaussian~\eqref{eq:gauss-switching}, 
normalised so that 
$\int_{-\infty}^{\infty} \chi^2(\tau) \, \text{d}\tau = 1$. 
Then 
\begin{align}
\hat\chi(\omega)  = 
\pi^{1/4}(2\sigma)^{1/2} \, 
e^{-\sigma^2\omega^2/2}
\ , 
\end{align}
and from \eqref{eq:bothmodes-response} we have 
\begin{subequations}
\label{eq:bothmodes-response-gaussian}
\begin{align}
F_{\text{osc}}(\Omega)
&= 
\frac{2 \pi^{3/2}\,\sigma}{L^2}
\sum_{\eta=\pm1}
\sum_{n=1}^{\infty}
e^{-2\eta\beta} n \, 
e^{
- \sigma^2 ( \Omega + 2\pi n e^{-\eta\beta}/L )^2}
\ , 
\label{eq:oscmode-response-gaussian}
\\
F_{\text{zm}}(\Omega)
&= 
\frac{2\pi^{1/2} \cosh^2 \! \beta}{L^2}
\langle\psi|
P_S^2
|\psi\rangle
\, 
\sigma e^{-\sigma^2\Omega^2}
\ .  
\label{eq:zmode-response-gaussian}
\end{align}
\end{subequations}

Consider first the limit of long detection, $\sigma\to\infty$, with $\beta$ fixed. 
In this limit \eqref{eq:bothmodes-response-gaussian} reduces to 
\begin{subequations}
\label{eq:bothmodes-transrate}
\begin{align}
F_{\text{osc}}(\Omega)
&= 
\frac{2\pi^2}{L^2}
\sum_{\eta=\pm1}
\sum_{n=1}^{\infty}
n e^{-2\eta\beta}
\, 
\delta \! \left(\Omega + \frac{2\pi n e^{-\eta\beta}}{L}\right)
\notag
\\
&= 
- \frac{\pi}{L}
\sum_{\eta=\pm1}
\sum_{n=1}^{\infty}
e^{-\eta\beta} \Omega 
\, 
\delta \! \left(\Omega + \frac{2\pi n e^{-\eta\beta}}{L}\right)
\label{eq:oscmode-transrate}
\ , 
\\
F_{\text{zm}}(\Omega)
&= 
\frac{2\pi \cosh^2 \! \beta }{L^2}
\langle\psi|
P_S^2
|\psi\rangle
\, 
\delta(\Omega)
\ , 
\label{eq:zmode-transrate}
\end{align}
\end{subequations}
where $\delta$ is Dirac's delta-function. $F_{\text{osc}}$
consists of strict delta-peaks of de-excitation, 
at the Doppler-shifted frequencies of the oscillator modes. 
$F_{\text{zm}}$ is a strict delta-peak at zero energy. 
Given that the detector's energy gap is assumed nonvanishing, 
the zero mode does not contribute to the response 
in the long detection limit. 

It can be verified that formulas \eqref{eq:bothmodes-transrate} also ensue if, 
instead of working with a switching function, 
we appeal to stationarity at the outset and consider the transition rate, 
\begin{align}
\dot{F}(\Omega)
= 
\int_{-\infty}^\infty 
\text{d}\tau 
\, 
e^{-\ii\Omega (\tau-\tau')}
\partial_\tau \partial_{\tau'}
W(\tau,\tau')
\ , 
\label{eq:stationary-transrate}
\end{align}
obtained from 
\eqref{eq:respfunction-der-def}
by setting $\chi(\tau)=1$ and formally factoring out the 
infinite total detection time \cite{Birrell1984}. 
This shows that the normalisation of our switching 
function \eqref{eq:gauss-switching} is well adapted for
recovering a transition rate per unit time in the 
$\sigma\to\infty$ limit. 

Consider then the limit of ultrarelativistic velocity, 
$|\beta|\to\infty$, with $\sigma$ fixed. 
$F_{\text{zm}}$ \eqref{eq:zmode-response-gaussian} 
diverges proportionally to $e^{2|\beta|}$. 
In $F_{\text{osc}}$~\eqref{eq:oscmode-response-gaussian}, 
the contribution from the blueshifted modes goes to zero, but 
an integral estimate shows that the contribution from the redshifted modes has 
a finite limit, given by 
\begin{align}
F_{\text{osc}}^{|\beta|\to\infty}(\Omega)
& = 
\frac{2 \pi^{3/2}\,\sigma}{L^2}
\int_0^\infty \text{d}x \, x 
\exp \! \left[
- \sigma^2 \left(
\Omega + \frac{2\pi x}{L}
\right)^2
\right]
\notag
\\[1ex]
& = 
\frac{1}{4 \sigma}
\left[
\frac{e^{-\sigma^2\Omega^2}}{\pi^{1/2}} - \sigma\Omega \erfc(\sigma\Omega)
\right]
\ . 
\label{eq:rel-vel-oscresp}
\end{align}
It can be verified, using the Minkowski vacuum Wightman function \eqref{eq:mink-wightman} 
and~\eqref{eq:respfunction-der-def},  
that \eqref{eq:rel-vel-oscresp} is exactly half of
the response of an inertial detector in the Minkowski vacuum in 
full Minkowski space. The physical picture of the ultrarelativistic motion is hence 
that the redshifted oscillator modes contribute to the response as if the spacetime 
were not periodic, while the contribution from the blueshifted oscillator 
modes is blueshifted beyond the energies accessible to the detector. 

We shall not attempt to take the limits of long detection 
and ultrarelativistic velocity simultaneously. However, if the zero-mode contribution
is considered negligible, we note that taking the two limits in succession commutes: 
both the $|\beta|\to\infty$ limit of \eqref{eq:oscmode-transrate} and the $\sigma\to\infty$ 
limit of \eqref{eq:rel-vel-oscresp} give 
\begin{align}
F(\Omega) = - \tfrac12 \Omega \Theta(-\Omega)
\ , 
\end{align}
where $\Theta$ is the Heaviside function. 
This is exactly half of the transition rate in inertial motion 
in Minkowski space in Minkowski vacuum, 
obtained from \eqref{eq:stationary-transrate} with~\eqref{eq:mink-wightman}. 

It is remarkable, and perhaps surprising, that the 
oscillator-mode contribution to the response 
exhibits no pathology in the ultrarelativistic limit. 
We see this as evidence that the periodic cavity provides a 
useful arena for analysing detector-field 
interaction even at relativistic velocities \cite{Brown2012,Fuenetesevolution,Wilson}.

\subsection{Uniformly accelerated detector}\label{subsec:unifacc}

As a second example, we consider a 
detector on the uniformly accelerated worldline  
\begin{align}
t = a^{-1}\sinh(a\tau)
\ , \ \ \ 
x = a^{-1}\cosh(a\tau)
\ , 
\label{eq:rindler-motion}
\end{align}
where $\tau$ is the proper time and the positive parameter $a$ is the proper acceleration. 
The detector accelerates towards increasing~$x$, and the detector is momentarily 
at rest in the cylinder's rest frame at the moment $\tau=0$. 

The trajectory is locally stationary with respect 
to the local Minkowski geometry in its neighbourhood, but it is 
not stationary with respect to the global time translations that leave the cylinder invariant. 
In geometric terms, the trajectory is invariant under the local boost-generating 
Killing vector $x\partial_t + t\partial_x$, but this Killing vector is not 
globally defined on the cylinder because of the spatial periodicity. 

Pulling back \eqref{eq:oscmode-wightman} and \eqref{eq:zmode-wightman}
to the detector's worldline, we find
\begin{widetext}
\begin{subequations}
\label{eq:bothmodes-corr-acc-dd}
\begin{align}
\partial_\tau \partial_{\tau'}
W_{\text{osc}}(\tau,\tau') 
&= 
\frac{\pi}{L^2}
\sum_{\eta=\pm1}
\sum_{n=1}^{\infty}
n\, e^{-\eta a (\tau+\tau')}
\exp \! \left[
\ii\eta \frac{2\pi n }{aL} (e^{-\eta a\tau} - e^{-\eta a\tau'} + \ii\eta\epsilon)
\right]
\label{eq:oscmode-corr-acc-dd}
\ , 
\\
\partial_\tau \partial_{\tau'}
W_{\text{zm}}(\tau,\tau') 
&= 
\frac{1}{L^2}
\langle\psi|
P_S^2
|\psi\rangle
\cosh(a\tau) \cosh(a\tau')
\ , 
\label{eq:zmode-corr-acc-dd}
\end{align}
\end{subequations}
where again $\eta=1$ comes from the 
right-movers and $\eta=-1$ comes from the left-movers. 
From \eqref{eq:respfunction-der-def} we then obtain 
\end{widetext}
\begin{subequations}
\label{eq:bothmodes--acc-response}
\begin{align}
F_{\text{osc}}(\Omega)
&= 
\sum_{\eta=\pm1}
\sum_{n=1}^{\infty}
\left|J_n^\eta\right|^2
\ , 
\\[1ex]
J_n^\eta
&= 
\frac{\sqrt{\pi n}}{L}
\int_{-\infty}^\infty
\text{d}\tau 
\, 
\chi(\tau) 
\, 
e^{-\ii\Omega\tau -\eta a \tau}
\notag
\\
& \hspace{13ex}
\times 
\exp \! \left(
\ii\eta \frac{2\pi n }{aL} e^{-\eta a\tau}
\right)
\ , 
\\[1ex]
F_{\text{zm}}(\Omega)
&= 
\frac{\langle\psi|
P_S^2
|\psi\rangle}{L^2}
\left|
\int_{-\infty}^\infty
\text{d}\tau 
\, 
\chi(\tau) 
\, 
e^{-\ii\Omega\tau}
\cosh(a\tau)
\right|^2
\ . 
\label{eq:zmode-acc-response}
\end{align}
\end{subequations}

As the trajectory is not stationary on the cylinder, 
we now consider the Gaussian switching function 
\begin{align}
\chi_{\tau_0}(\tau)  = 
\frac{1}{\pi^{1/4}\sigma^{1/2}}
e^{- (\tau-\tau_0)^2/(2\sigma^2)}
\ , 
\label{eq:gauss-shift-switching}
\end{align}
where $\sigma>0$ as before but the new real-valued parameter $\tau_0$ 
specifies the instant about which $\chi_{\tau_0}$ is peaked. 
From \eqref{eq:bothmodes--acc-response}, we then have 
\begin{subequations}
\label{eq:bothmodes--acc-gaussian-response}
\begin{align}
F_{\text{osc}}(\Omega)
&= 
\sum_{\eta=\pm1}
\sum_{n=1}^{\infty}
\left|J_n^\eta\right|^2
\ , 
\label{eq:oscmodes--acc-gaussian-response}
\\[1ex]
J_n^\eta
&= 
\frac{\pi^{1/4} n^{1/2}}{La \, \sigma^{1/2}}
e^{-(\eta a + \ii \Omega) \tau_0}
\int_0^\infty
\text{d}x
\, x^{\ii \eta \Omega/a}
\notag
\\[1ex]
& 
\hspace{4ex}
\times 
\exp \! \left[
- \frac{(\ln x)^2}{2 a^2 \sigma^2}
+ \ii \eta e^{-\eta a \tau_0}\frac{2\pi n }{aL} x
\right] 
\ , 
\\[1ex]
F_{\text{zm}}(\Omega)
&= 
\frac{2 \pi^{1/2} \sigma}{L^2}
\langle\psi|
P_S^2
|\psi\rangle
\, 
e^{-\sigma^2\Omega^2 +\sigma^2 a^2}
\notag
\\[1ex]
& \hspace{5ex}
\times \left[ \cos^2(\sigma^2 a \Omega)
+ 
\sinh^2(a\tau_0)
\right]
\ . 
\end{align}
\end{subequations}
Both the oscillator-mode contribution and the zero-mode contribution depend on~$\tau_0$. 

\begin{figure}[t]
\includegraphics[width=.43\textwidth]{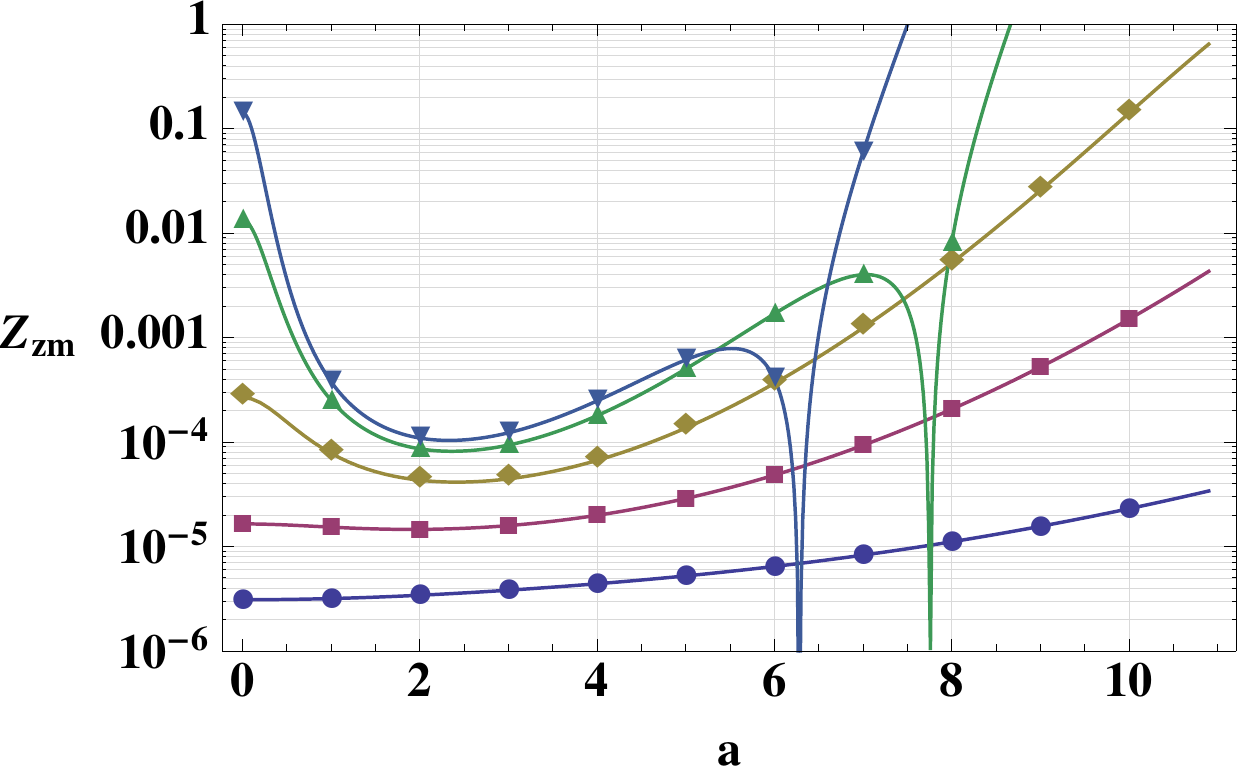}
\caption{(Color online): The plot shows $Z_{\text{zm}}:=F_\text{zm}/F_{\text{osc}}$ evaluated from 
\eqref{eq:bothmodes--acc-gaussian-response} with $L=1$, $\Omega=1$, $\tau_0=0$ 
and $\langle\psi| P_S^2 |\psi\rangle = 10^{-6}$, as a function of~$a$, with selected values of~$\sigma$: 
solid circle is $\sigma=0.15$, solid square is $\sigma=0.25$, rhombus is $\sigma=0.35$, 
triangle is $\sigma=0.45$, and inverted triangle is $\sigma=0.5$. 
$Z_{\text{zm}}$ has exact zeroes at $a= (\sigma^2\Omega)^{-1}(\pi/2+k\pi)$, $k = 0,1,\ldots$, 
of which only the first zero ($k=0$) for the largest two values of $\sigma$ is in the range covered by the plot.}
\label{estimator2}
\end{figure}

For the zero-mode contribution, we recall that $F_{\text{zm}}$ can be always made as 
small as desired by choosing $\ket{\psi}$ so that the overall coefficient $\langle\psi| P_S^2 |\psi\rangle$ is small. 
A more interesting question however is how the relative magnitudes of $F_{\text{zm}}$ 
and $F_{\text{osc}}$ depend on the other parameters when $\langle\psi| P_S^2 |\psi\rangle$ is fixed. 
We illustrate this in Fig.~\ref{estimator2}, showing a parameter range in which 
$F_{\text{zm}}/F_{\text{osc}}$ tends to grow with increasing 
interaction time and with increasing acceleration. 

\begin{figure}[t]
\includegraphics[width=.43\textwidth]{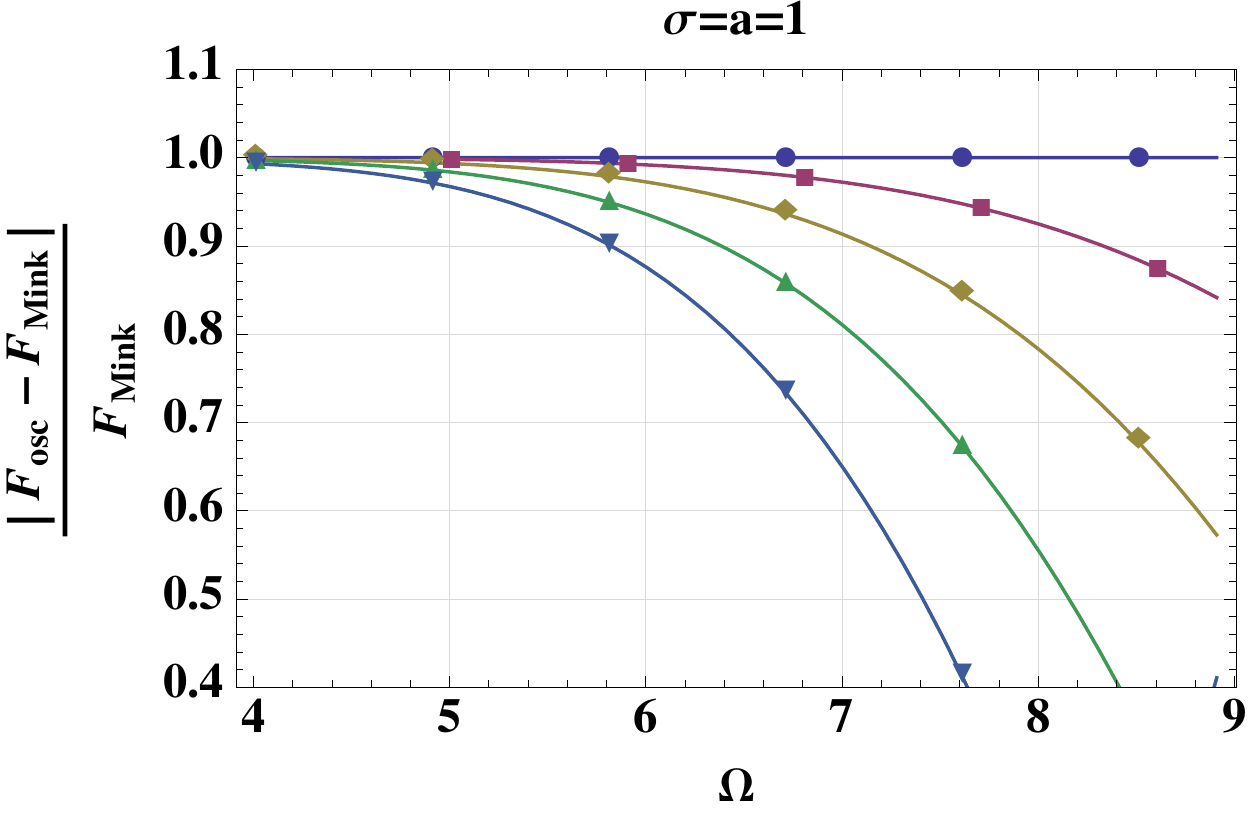}
\caption{(Color online): The plot shows $|F_\text{Mink}-F_\text{osc}| / F_\text{Mink}$, 
evaluated from 
\eqref{eq:bothmodes--acc-gaussian-response} and \eqref{eq:mink-rindler-response}, 
with $a = \sigma=1$ and $\tau_0=0$, as a function of~$\Omega$, 
with selected values of~$L$: 
solid circle is $L=0.01$, solid square is $L=0.15$, rhombus is $L=0.2$, 
triangle is $L=0.25$, and inverted triangle is $L=0.3$. 
$F_\text{osc}$ and $F_\text{Mink}$ get closer to each other when $L$ increases.
The slow convergence of the sum in 
\eqref{eq:oscmodes--acc-gaussian-response} 
limits our ability to go to higher values of~$L$.}
\label{comparL}
\end{figure}

For the oscillator-mode contribution, we expect that $F_{\text{osc}}$ should reduce to the 
response of the accelerating detector \eqref{eq:rindler-motion}
in Minkowski space in the Minkowski vacuum. 
The response in the Minkowski vacuum is given by 
\begin{align}
F_{\text{Mink}}(\Omega)
&= 
\frac{a e^{-\sigma^2 \Omega^2}}{4\pi}
\int_{-\infty}^\infty\!
\frac{\text{d}r}{\cosh^2 \! r}
\notag
\\[1ex]
& 
\hspace{3ex}
\times \exp{\left\{
- \frac{1}{a^2\sigma^2}
\left[
r + \ii \left(\sigma^2 a \Omega - \frac{\pi}{2}\right)
\right]^2
\right\}}
\ , 
\label{eq:mink-rindler-response}
\end{align}
as can be verified using 
\eqref{eq:mink-wightman}, \eqref{eq:respfunction-der-def}
and~\eqref{eq:rindler-motion}, and in the long detection time limit 
$F_{\text{Mink}}$ reduces to the 
familiar Planckian result in 
the Unruh temperature~$a/(2\pi)$, 
\begin{align}
F_{\text{Mink}}^{\sigma\to\infty}(\Omega)
= 
\frac{\Omega}{e^{2\pi\Omega/a} -1}
\ , 
\label{eq:planckian-response}
\end{align}
as can be verified using formula 3.985.1 in~\cite{grad-ryzh}. 
Numerical evidence for closeness of $F_{\text{osc}}$ and $F_{\text{Mink}}$ with increasing $L$ 
is given in Fig.~\ref{comparL}. The slow convergence of the sum in 
\eqref{eq:oscmodes--acc-gaussian-response} 
has prevented us from 
seeking evidence in a more extensive range of the parameter space. 

A limit of particular interest for the oscillator modes would be that of large $|\tau_0|$ 
with all other parameters fixed. In this limit the detector is 
moving through the cavity with an ultrarelativistic velocity 
throughout the vast majority of the effective detection time. 
Our ultrarelativistic inertial detector result \eqref{eq:rel-vel-oscresp}
suggests that in this limit $F_{\text{osc}}$ should tend to half 
of the Minkowski vacuum response 
$F_{\text{Mink}}$ \eqref{eq:mink-rindler-response}. 
If correct, this would be further evidence  
that the periodic cavity provides a 
useful arena for analysing detector-field 
interaction even at relativistic velocities \cite{Brown2012,Fuenetesevolution,Wilson}. 
The slow convergence of the sum in 
\eqref{eq:oscmodes--acc-gaussian-response} 
has however not enabled us to obtain conclusive evidence about this question. 

$F_{\text{osc}}$ can be written 
in a form that avoids a discrete sum by using for the Fock vacuum Wightman function the summed expression 
\eqref{eq:oscmode-wightman-summed}, and the $\epsilon$-regulator can then be eliminated by the techniques of 
\cite{BenitoJorma,Satz2006,Satz2007,Louko:2007mu,Hodgkinsonclick,Hodgkinson2012}. 
We have not investigated whether this integral representation 
improves the stability of numerical evaluation in the limits of interest.

\section{Conclusions\label{sec:conclusions}}

We have assessed  the impact of the zero mode of a quantum field on the dynamics of particle detectors. This mode is often neglected when considering the light-matter interaction under periodic or Neumann  boundary conditions.

We worked with a massless scalar field in a periodic cavity in $(1+1)$-dimensional Minkowski space. 
We considered the traditional Unruh-DeWitt (UDW) detector, coupled linearly to the field, as well as a modified UDW detector that 
couples linearly to the proper time derivative of the field. We treated the interaction perturbatively, to quadratic order in the coupling constant for the detector's transition probabilities. 

For the UDW detector, we first showed that when the oscillator modes of the field are initially in a Fock state, or in an ensemble of Fock states, the zero mode of the field is not affected by the backreaction of the non-zero modes on the detector, or vice versa. To quadratic order in the coupling constant, there is hence no danger that energy in the oscillator modes of the field would get transferred to the zero mode via the interaction through the detector, regardless of the state of the detector. This conclusion does however not need to hold if the oscillator modes of the field are initially in a state with a non-diagonal density matrix in the Fock basis, such as a coherent state. 

We then showed that the zero mode does have a non-vanishing 
direct effect on the evolution of the detector's density matrix. 
This effect can be made negligible in standard quantum optical settings, 
but situations in which the effect can be significant, and even dominant, 
can arise in other settings, including the Unruh effect \cite{Unruh1} 
or the harvesting of quantum entanglement from a field~\cite{Reznik1}. 

For the derivative-coupling detector, 
we found that the zero mode has again a nonvanishing direct effect on the 
detector's transition probabilities, but this effect can be made 
as small as desired by just tuning the detector's initial state. 
The effect is invariant under time translations in the cavity, and it is directly 
proportional to the contribution of the zero mode to the field's renormalised 
stress-energy tensor. As examples, we considered a detector moving 
inertially but with an arbitrary velocity, including the ultrarelativistic limit, 
and a detector in uniformly accelerated motion. 

Our analysis provides the basic tools for studying 
the Unruh effect in a periodic cavity for the derivative-coupling detector. 
Exploiting these tools for a systematic survey of the parameter space, 
whether by analytic or numerical techniques, is left to future work. 

Systems where a zero mode arises as a consequence of Neumann boundary conditions will have some quantitative differences because of the absence of spatial homogeneity, but we anticipate the qualitative conclusions about the zero mode to be largely similar. We also anticipate that both our analysis and our conclusions can be generalised to cosmological spacetimes in which zero modes arise~\cite{QuanG}.

 It should be interesting to study in detail possible zero-mode effects on relevant well-known results in relativistic quantum information such as, for instance, vacuum entanglement harvesting and farming \cite{Reznik1,farming} and relativistic quantum communication \cite{Robort,Qcollcall}, when considered in the context of periodic and Neumann cavities. Since we showed that in many relevant cases there is no backreaction of the zero-mode on the oscillatory modes through their interaction with the detector, it is doubtful that the zero mode dynamics alone could destabilize the process of entanglement harvesting or hinder relativistic communication protocols. Nevertheless, as shown by the full density matrix analysis carried over in Section \ref{subsec:solution}, there would be an impact on these phenomena coming from the dynamics of the zero mode that might become relevant in some regimes. Although outside of the scope of this paper, it may be relevant to study the role of the zero-mode dynamics in those scenarios in future work. 

As a final comment, we note that while the focus of this paper is theoretical, 
the conclusions are applicable in 
all detector-field interaction settings where there are periodic or Neumann boundary conditions. 
Among laboratory systems this includes closed optical cavities, 
such as optical-fibre loops \cite{Tsuchida:90}, 
and superconducting circuits coupled to periodic \cite{Ultrastrong} 
or Neumann microwave guides~\cite{DCasimir}.

\section*{Acknowledgments}

We thank 
William Donnelly, 
Robert Jonsson, 
Benito Ju\'arez-Aubry, 
Achim Kempf,  
Don Marolf 
and 
Adrian Ottewill 
for helpful discussions and correspondence. 
Both authors thank
Achim Kempf, Robert Mann and Gerard Milburn for hospitality at the 
Banff International Research Centre 
workshop 13w5153 Entanglement in Curved Spacetime, where this work was begun. 
J.L. thanks Achim Kempf and Eduardo Mart\'in-Mart\'inez 
for hospitality at the University of Waterloo. 
E.M-M. acknowledges the support of the Banting Postdoctoral Fellowship Programme. 
J.L. was supported in part by STFC (Theory
Consolidated Grant ST/J000388/1).

\bibliography{references}

\end{document}